\documentclass[iop]{emulateapj}

\usepackage{lineno}
\usepackage{color}
\usepackage{hyperref}  
\usepackage{breakurl}  

\newcommand\ms{\ifmmode{\rm m\thinspace s^{-1}}\else m\thinspace s$^{-1}$\fi}
\newcommand\kms{\ifmmode{\rm
 km\thinspace s^{-1}}\else km\thinspace s$^{-1}$\fi}

\newcommand{\actaa}{Acta Astronomica}
\newcommand{\pasa}{Publ.\ Astron\ Soc.\ of Australia}
\newcommand{\na}{New Astronomy}
\newcommand{\napp}{\vskip 3pt\noindent}

\shortauthors{Torres et al.}
\shorttitle{Runaway Stars}

\begin{document}
\submitted{Published in Monthly Notices of the Royal Astronomical Society, 539,
p.\ 282--296 (2025)}

\title{Spectroscopic Follow-Up of Candidate Runaway Stars}

\author{Guillermo Torres\altaffilmark{1},
Ralph Neuh\"auser\altaffilmark{2},
Sebastian A.\ H\"uttel\altaffilmark{2}, and
Valeri V.\ Hambaryan\altaffilmark{2,3,4}
}

\altaffiltext{1}{Center for Astrophysics $\vert$ Harvard \& Smithsonian,
  60 Garden St., Cambridge, MA 02138, USA; gtorres@cfa.harvard.edu}

\altaffiltext{2}{Astrophysical Institute, Friedrich-Schiller-University Jena, 
Schillerg\"a\ss chen 2, 07745 Jena, Germany}

\altaffiltext{3}{Byurakan Astrophysical Observatory after V.A.\ Ambartsumian, 0213, 
Byurakan, Aragatzotn, Armenia}

\altaffiltext{4}{Astrophysical Research Laboratory of Physics Institute, Yerevan 
State University, Armenia}

\begin{abstract}
Runaway stars are characterized by higher space velocities than typical
field stars. They are presumed to have been ejected from their birth
places by one or more energetic mechanisms, including supernova
explosions. Accurate radial velocities are essential for investigating
their origin, by tracing back their Galactic orbits to look for close
encounters in space and in time with neutron stars and young associations.
While most studies of runaways have focused
on OB stars, later-type stars have also been considered on occasion.
Here we report the results of a long-term high-resolution spectroscopic
monitoring program
with the goal of providing accurate radial velocities for
188 runaway candidates of spectral type A and later, proposed by
\cite{Tetzlaff:2011}.
We obtained multiple measurements
over a period of about 13~yr to guard against the possibility that some may be
members of binary or multiple systems, adding archival observations going back
another 25~yr in some cases.
We report new spectroscopic orbital solutions for more than
three dozen systems. Many more are also found to be binaries based on
available astrometric information. A small-scale study carried out here
to trace back the
paths of our targets together with those of four well-studied, optically-visible neutron stars among the so-called Magnificent Seven, resulted in no credible encounters.
\end{abstract}


\section{Introduction}
\label{sec:introduction}

The peculiar velocities of stars represent their motion with respect to their
surroundings or to their birth cluster.  While the
dispersion of the peculiar 3D velocities of most stars is only a few
$\kms$, some show much higher velocities, presumably because they have been
ejected from their birthplaces. They are referred to as `runaway' stars.
They occupy a Maxwellian tail in the distribution of peculiar
velocities, which is distinguishable from the larger and relatively
narrow peak populated by normal field stars \citep[see, e.g.,][Figure~1]{Tetzlaff:2011}. The dividing line, as proposed in that
study and others, is usually taken to be around 30~\kms.
Numerical simulations \citep[e.g.,][]{Renzo:2019} show that,
depending on initial conditions, stars can
also be ejected with lower velocities, such as from a binary with a
wide orbit. Such objects have been called `walkaways'.

Several mechanisms have been suggested for a star to attain higher
than normal velocities: (i) Dynamical ejection from the birth cluster
or association within the first few Myr, through interactions
among three or more stars in the core of a dense cluster \citep{Poveda:1967}; (ii)
Binary supernova ejection, in which the former companion of a massive
star exploding as a supernova can become unbound, if the ejected mass
and/or the kick of the newborn neutron star is sufficiently high
\citep{Blaauw:1961}. Ejection velocities from this and the previous
mechanism can be as high as 100~\kms; (iii) Black hole ejection, in
which stars can be accelerated to even higher (possibly escape) velocities by close
encounters with the supermassive black hole at the centre of our Galaxy
\citep{Hills:1988}. These ejected objects are referred to as
hypervelocity stars \citep{Brown:2015}.

Aside from the interesting question about their origin, stars ejected from
their birthplaces are also an important component to include in the calculation
of present-day mass functions, when comparing them to
initial mass functions (IMFs).

Runaway stars are usually investigated by tracing back their 3D motion
through the Galactic potential, to find instances in
which, e.g., an OB star and a neutron star may have been at the same
location in space at the same time. This could be evidence for a
supernova occurring in a binary, ejecting both the pulsar and the
runaway star \citep[see, e.g.,][]{Neuhauser:2020}. In practice, the validity of
traceback calculations is typically limited to a few tens of Myr due
to various observational uncertainties. Thus, the ejection is required
to have happened within that time frame. For this reason, OB stars are
most favourable for this type of analysis, because their short
main-sequence lifetimes imply that they must be young (less than
$\sim$30~Myr for spectral types earlier than B3). Nevertheless,
later-type stars can also be ejected as runaway stars by any of the
scenarios described above. Unfortunately, however, it may be more
difficult to precisely determine their ages, by methods such as the
use of the Li~$\lambda$6708 absorption line as a youth diagnostic, or by
other means.
For example, lithium in stars is destroyed at temperatures above about $2.5 \times 10^6$~K as it gets mixed into the interior \citep[e.g.,][]{Bodenheimer:1965, Neuhauser:1997}, but the rate of Li depletion for mid-F to late-M dwarfs is influenced by the extent of the convection zone. It has a complex and not fully understood dependence on temperature, mass, and metallicity, and may also depend on rotation and stellar activity \citep[e.g.,][]{Soderblom:1993}. On the other hand, age determinations using stellar evolution models are limited in their sensitivity, because observables such as luminosity, temperature, and surface gravity change very slowly with time upon
arrival of the star on the main sequence. Other techniques such as asteroseismology are not widely applicable for most field stars, due to a
lack of suitable observations.
Late-type runaway stars have therefore received less attention than
early-type stars, but are no less important.
They are the subject of this paper.

An extensive catalogue of candidate runaway
stars of all spectral types within 3~kpc of the Sun was published some
years ago by \cite{Tetzlaff:2011}. Those objects were presumed to be
young based on the application of several criteria, but they were
considered merely as candidates because their radial velocities (RVs)
were largely unknown at
the time. Their proper motions and parallaxes relied on the Hipparcos
catalogue \citep{vanLeeuwen:2007}.  Particularly with its most recent data release (DR3), the Gaia mission
\citep{Gaia:2023} has now provided greatly improved parallaxes and proper
motions for large numbers of stars, as well as average radial
velocities in many cases. However, the cadence of the spectroscopic
observations is
such that spectroscopic binaries or higher-order multiples among the
runaway candidates may be missed, or their orbits may not be solvable
with the Gaia data alone. This can lead to mean velocities that are biased.  While
it may be true that most runaway stars are expected to be single on
account of the violence
of the ejection mechanism, close binaries are possible as well, and in fact
appear to be quite common among walkaway stars \citep[e.g.,][]{Bhat:2022}.
Therefore, repeated ground-based RV measurements remain important for
the study of runaway stars, in order to characterize them fully and
accurately. 

With that in mind, soon after the \cite{Tetzlaff:2011} list
appeared, we embarked on a long-term project to monitor the RVs of a
subset of about 190 of them with spectral types A and later. The main
goal was to provide precise
and accurate RVs, accounting for binarity for objects found to be in
multiple systems.  Here we report on the full observational results of that survey,
including the discovery of many binaries among our targets, of which
Gaia has detected only a small fraction.
We also carry out a pilot study to explore the possible origin of these
stars, in the context of the supernova ejection scenario described earlier.

Our paper begins with an explanation of how the list of targets we followed
up was assembled (Section~\ref{sec:sample}), and continues in Section~\ref{sec:spectroscopy} with a description
of our spectroscopic material and the derivation of radial velocities for
single- and double-lined objects. We then comment on how we identified
binaries among our targets (Section~\ref{sec:binaries}),
based both on the RVs and on astrometric information
from various sources. This section also presents the spectroscopic
orbits we have determined. The distribution of peculiar velocities
for our sample is reported in Section~\ref{sec:spacevelocities}. Then,
as a preview of a fuller study left for a
future paper, Section~\ref{sec:traceback} presents a small-scale analysis in
which we trace back the 3D motion of our targets together with a handful of
well-studied neutron stars, to explore a possible connection consistent
with the supernova ejection scenario. We end with final remarks in Section~\ref{sec:conclusions}.
An Appendix is included in which we collect
the orbital determinations from Gaia for comparison with ours, and 
provide details of interest for selected targets.

\section{The Sample}
\label{sec:sample}

The target list for our program, provided by N.\ Tetzlaff (2013,
priv.\ comm.), was selected mostly from the catalogue of candidate runaway stars
of \cite{Tetzlaff:2011}, subject to the condition that the stars be located
north of about declination $-35\arcdeg$ in order to be observable with the
telescope facilities at our disposal. They were chosen for
spectroscopic follow-up because, at the time of that study, they
lacked a radial-velocity measurement, and only the tangential
velocities could be computed based on the Hipparcos proper motions and
parallaxes.  A total of 159 objects were taken from those authors'
Table~4, which lists candidates in which the runaway probability was
estimated to be higher than 50 per cent in at least one of the peculiar
tangential velocity components.  An additional 15 stars come from
Table~1 of \cite{Tetzlaff:2011}, which includes objects with poorly
determined kinematics that they deemed to be young, a condition
required for runaway status. The remaining 14 stars were added based
on information gathered after the 2011 publication. 

The complete list of 188 targets, ranging from main-sequence stars
to supergiants and with spectral types A through M, is given in
Table~\ref{tab:sample}, with information
collected from the Gaia DR3 catalogue. In addition to the parallaxes,
proper motions, and $G$-band magnitudes, we include the zero-point corrections
to the parallaxes ($\Delta\pi$) advocated by \cite{Lindegren:2021}, as
well as the Renormalised Unit Weight Error (RUWE), which is an indicator
of the quality of the astrometric fit, and possibly an indicator of
binarity (see below). Spectral types were taken from
the catalogue of \cite{Tetzlaff:2011}, or from SIMBAD.

\setlength{\tabcolsep}{3pt}
\begin{deluxetable*}{lccccccccc}
\tablewidth{0pc}
\tablecaption{Sample of Targets. \label{tab:sample}}
\tablehead{
\colhead{Name} &
\colhead{Gaia ID} &
\colhead{$\pi_{\rm DR3}$} &
\colhead{$\Delta\pi$} &
\colhead{$\mu_{\alpha}\cos\delta$} &
\colhead{$\mu_{\delta}$} &
\colhead{RUWE} &
\colhead{$G$} &
\colhead{SpT} &
\colhead{$V_{\rm LSR}$}
\\
\colhead{} &
\colhead{} &
\colhead{(mas)} &
\colhead{(mas)} &
\colhead{(mas yr$^{-1}$)} &
\colhead{(mas yr$^{-1}$)} &
\colhead{} &
\colhead{(mag)} &
\colhead{} &
\colhead{(km s$^{-1}$)}
}
\startdata
%
HIP001008 &   385608215446223488 &  $1.6993 \pm 0.0525$      & 0.0305  &  $-8.819 \pm 0.040$         &  \phn$-6.195 \pm 0.033$     &  2.557  &  6.877 & K0 &  $44.07 \pm 1.10$ \\
HIP001479 &   567394492955490432 &  $1.5054 \pm 0.0143$      & 0.0245  &  \phs$26.513 \pm 0.016$\phn &  \phn\phs$7.288 \pm 0.020$  &  0.897  &  7.414 & K5 &  $70.83 \pm 1.02$ \\   
HIP001602 &  2747597682852427776 &  $2.8742 \pm 0.2077$      & 0.0355  &  $-2.451 \pm 0.214$         &  \phs$13.403 \pm 0.155$     &  7.260  &  6.240 & K0 &  $49.79 \pm 1.18$ \\   
HIP001733 &   537077349608343808 &  $3.9668 \pm 0.0716$      & 0.0188  &  $-6.427 \pm 0.073$         &  \phn$-0.531 \pm 0.074$     &  5.315  &  8.719 & A2 &  $24.59 \pm 0.68$ \\  
HIP002710 &  2526580040189355136 &  $24.7805 \pm 0.0270$\phn & 0.0358  &  $-71.190 \pm 0.035$\phn    &  $-85.398 \pm 0.025$        &  1.096  &  6.801 & F2 &  $32.44 \pm 0.70$   
\enddata
\tablecomments{Following the target name and Gaia ID, subsequent columns contain
the Gaia DR3 parallax ($\pi_{\rm DR3}$), an additive adjustment ($\Delta\pi$) to the
parallax to correct for a zero-point offset known to exist in the
Gaia DR3 catalogue \citep[see][]{Lindegren:2021}, the proper motion
components ($\mu_{\alpha}\cos\delta$, $\mu_{\delta}$), the 
Renormalised Unit Weight Error (RUWE) from Gaia DR3, the Gaia magnitude
($G$), the spectral type (SpT) from \cite{Tetzlaff:2011} or SIMBAD,
and the velocity of each target relative to the Local
Standard of Rest (see Section~\ref{sec:spacevelocities}). The latter velocities
were computed from the other columns and the mean radial velocities reported below,
  with the solar motion being adopted from \cite{Schonrich:2010}. For HIP056703,
  the parallax and proper motion components are taken from the Gaia DR2
  catalogue \citep{Gaia:2018}, as DR3 does not list those values for this star.
  For the few targets brighter than $G = 6$ that are
  beyond the range of applicability of the parallax adjustments
  by \cite{Lindegren:2021}, the correction $\Delta\pi$ is approximate.
  (This table is available in its entirety in machine-readable form.)}
\end{deluxetable*}
\setlength{\tabcolsep}{6pt}

\section{Spectroscopic Observations}
\label{sec:spectroscopy}

Observations of our targets were conducted at the Center for
Astrophysics (CfA) using several telescope/instrument combinations.
Most of the spectra, beginning in September of 2009, were collected
with the Tillinghast Reflector Echelle Spectrograph
\citep[TRES;][]{Furesz:2008, Szentgyorgyi:2007}
on the 1.5-m Tillinghast reflector at the Fred L.\ Whipple Observatory,
on Mount Hopkins (AZ, USA). This is a
bench-mounted, fibre-fed instrument delivering a resolving power of $R
\approx 44,\!000$ in 51 orders, over a wavelength range of
3800--9100~\AA. The signal-to-noise ratios of the 1898 spectra  we
obtained depend strongly on the brightness and sky conditions, and
range from about 10 to 400 per resolution element of 6.8~\kms.
Reductions were performed with a dedicated pipeline
\citep[see][]{Buchhave:2012}. The velocity zero-point was monitored
with observations of IAU standards each run, and appropriate
corrections were applied to remove any drifts with time.

Earlier archival observations of a few stars were gathered with two
nearly identical spectrographs, attached to the Tillinghast reflector
and to the 1.5-m Wyeth reflector at the Oak Ridge Observatory (now
closed), in the town of Harvard (MA, USA). These instruments \citep[Digital
  Speedometers;][]{Latham:1992} were equipped with intensified photon-counting
Reticon detectors that limited the recorded output to a single echelle
order about 45~\AA\ wide, centred on the \ion{Mg}{1}\,b triplet near
5187~\AA. They delivered a resolving power of $R \approx 35,\!000$.
Signal-to-noise ratios for these observations ranged between 10 and 60
per resolution element of 8.5~\kms. Reductions were carried out with a
dedicated pipeline \citep{Latham:1985, Latham:1992}, and the velocity
zero-point was monitored with observations of the twilight sky at dusk
and dawn. As with TRES, run-to-run corrections were applied to
maintain a stable velocity system.
A total of 181 spectra were obtained with these instruments.

The native velocity system of the Digital Speedometers is slightly
offset from the IAU system by 0.14~\kms\ \citep{Stefanik:1999}, as
determined from observations of minor planets in the solar system. In
order to remove this shift, we added a correction of +0.14~\kms\ to
all our raw velocities from these instruments. Similarly, observations
of asteroids were used to place the measurements from TRES also on the
IAU system.

Radial velocities for single-lined objects were determined by
cross-correlation, using templates from a large pre-computed library of
synthetic spectra based on model atmospheres by R.\ L.\ Kurucz
\citep[see][]{Nordstrom:1994, Latham:2002}, coupled with a line list tuned to
better match real stars. The templates are restricted to a region near
the \ion{Mg}{1}\,b triplet, which captures most of the velocity
information. The template parameters were chosen as follows. For
simplicity, we adopted the working assumption of solar composition
for all our targets, as this has little effect on the velocities. 
Surface gravity ($\log g$) is a subtle effect
that is challenging to determine accurately from our spectra, in part
because it is strongly correlated with the effective temperature
($T_{\rm eff}$). As $\log g$ also has little impact on the velocities, we set
it by locating each target in the colour-magnitude diagram, based on
their $G_{\rm BP}-G_{\rm RP}$ colour index from Gaia and their absolute
$G$-band magnitude $M_G$. We used stellar evolution models from the
PARSEC~v1.2S series \citep{Chen:2014} to infer
a rough surface gravity, and then rounded off these $\log g$ values to
the nearest step in our grid of templates (step size =
0.5~dex). Effective temperatures and suitable rotational
broadenings\footnote{Strictly speaking, the rotational broadening we
  infer for each star includes a contribution from the difference
  between the true macroturbulent velocity and the value of that
  parameter that is built into our
  templates ($\zeta_{\rm RT} = 1~\kms$). For simplicity, we refer
  to the line broadening here as $v \sin i$.} for the templates ($v \sin
i$) were chosen following the procedure of \cite{Torres:2002}, in
which grids of correlations were run over wide ranges in those
parameters, to identify the combination giving the highest correlation
value averaged over all exposures.

In a few cases the model-inferred $\log g$ values were beyond the
range of our template library, or the temperatures reached the upper
or lower limits available. For those targets, we manually adjusted
either $\log g$ or $T_{\rm eff}$ to bring them within range, and reran
the procedure of \cite{Torres:2002} to maximize the average
correlation, thereby minimizing any possible bias in the velocities.

For objects whose spectra are double-lined, RVs were measured using
TODCOR \citep{Zucker:1994}, a two-dimensional cross-correlation
algorithm that uses two templates matched to the components. To set
the parameters of the synthetic spectra for these cases, we applied a similar
methodology as explained above, successively optimizing the primary
and secondary templates by iterations. We also measured the flux
ratio between the components at the mean wavelength of our observations.

The individual RVs with their formal (internal) uncertainties for the
single-lined objects are listed in Table~\ref{tab:rvs1}. Those for the
double-lined targets are given in Table~\ref{tab:rvs2}.

\setlength{\tabcolsep}{8pt}
\begin{deluxetable}{lccc}
\tablewidth{0pc}
\tablecaption{Radial Velocities for the Single-lined Objects. \label{tab:rvs1}}
\tablehead{
\colhead{Name} &
\colhead{BJD} &
\colhead{RV} &
\colhead{$\sigma$}
\\
\colhead{} &
\colhead{(2,400,000+)} &
\colhead{(\kms)} &
\colhead{(\kms)}
}
\startdata
HIP001008  &  56550.8070  &  $-$19.13  &  0.15 \\
HIP001008  &  56585.7152  &  $-$18.03  &  0.15 \\
HIP001008  &  56606.6593  &  $-$17.47  &  0.15 \\
HIP001008  &  56638.6765  &  $-$16.80  &  0.16 \\
HIP001008  &  56652.6610  &  $-$16.67  &  0.16 
\enddata
\tablecomments{(This table is available in its entirety in machine-readable form.)}
\end{deluxetable}
\setlength{\tabcolsep}{6pt}

\setlength{\tabcolsep}{2pt}
\begin{deluxetable}{lccccc}
\tablewidth{0pc}
\tablecaption{Radial Velocities for the Double-lined Objects. \label{tab:rvs2}}
\tablehead{
\colhead{Name} &
\colhead{BJD} &
\colhead{RV$_1$} &
\colhead{RV$_2$} &
\colhead{$\sigma_1$} &
\colhead{$\sigma_2$}
\\
\colhead{} &
\colhead{(2,400,000+)} &
\colhead{(\kms)} &
\colhead{(\kms)} &
\colhead{(\kms)} &
\colhead{(\kms)}
}
\startdata
HIP012297  &  56554.8669  &  \phn$-$1.95  &  $-$21.94      &  0.60  &  \phn4.57  \\
HIP012297  &  56588.9048  &  \phn$-$8.86  &  \phn$-$8.90   &  0.63  &  \phn4.81  \\
HIP012297  &  56609.8327  &  $-$18.65     &  \phs\phn4.43  &  1.88  &  14.29     \\
HIP012297  &  56641.6258  &  $-$27.13     &  \phs17.42     &  1.05  &  \phn8.00  \\
HIP012297  &  56652.6923  &  $-$20.66     &  \phs19.48     &  0.79  &  \phn6.03 
\enddata
\tablecomments{(This table is available in its entirety in machine-readable form.)}
\end{deluxetable}
\setlength{\tabcolsep}{6pt}

\section{Binaries in the Sample}
\label{sec:binaries}

A summary of the velocity information for each target based on
the measurements obtained here is presented in
Table~\ref{tab:rvsummary}, and includes the time span and number of
observations, as well as the weighted mean velocity and corresponding
uncertainty. These mean velocities are used below 
for computing the Galactic orbits of the stars, in combination with the
proper motions and parallaxes listed earlier. A large number of objects
in our sample are binary or multiple systems of one type or another.
In these cases, the mean velocities may not be representative of the
true line-of-sight motion. We discuss these situations below.

\setlength{\tabcolsep}{4pt}
\begin{deluxetable}{lccccc}
\tablewidth{0pc}
\tablecaption{Summary of Radial-Velocity Information. \label{tab:rvsummary}}
\tablehead{
\colhead{Name} &
\colhead{Time Span} &
\colhead{$N_{\rm obs}$} &
\colhead{Mean RV} &
\colhead{Error} &
\colhead{Notes}
\\
\colhead{} &
\colhead{(days)} &
\colhead{} &
\colhead{(\kms)} &
\colhead{(\kms)} &
\colhead{}
}
\startdata
  HIP001008  &  1448.1         &  25  &  $-$16.77     &  0.04  &  SB1  \\
  HIP001479  &  \phm{22}75.8   &   3  &  $-$13.54     &  0.08  &       \\
  HIP001602  &  1390.2         &  20  &  \phs36.21    &  0.02  &  SB1  \\
  HIP001733  &  4087.9         &  22  &  \phn$-$2.93  &  0.72  &  VAR  \\
  HIP002710  &  9951.8         &  10  &  \phs12.32    &  0.12  &       
\enddata

\tablecomments{Objects flagged in the notes as `SB1' or `SB2' are
  single- or double-lined binaries for which we have derived an orbit.
  We use `SB1[GAIA]' for cases where our own RVs show little or no change,
  whereas Gaia has reported an orbit. The `VAR' and `VAR?'
  classifications indicate certain or probable RV variability, which
  in most cases is a sign of binarity, but can also be due to
  pulsation.  The `VIS' code flags objects having
  visual companions with confirmed or possible physical association,
  and those discovered from their astrometric motion by Gaia DR3
  are indicated with `AST[GAIA]'. Objects showing acceleration
  in the plane of the sky from Hipparcos and Gaia data \citep{Brandt:2021} have the code `ACC',
  and those with Gaia RUWE values larger than 1.4, suggestive of binarity,
  are flagged with `RUWE'.
  For spectroscopic binaries with orbits (from our own measurements,
  or from Gaia), the mean RV listed is the centre-of-mass velocity.
  For HIP017878, which is a W~UMa overcontact system, we adopted the
  $\gamma$ velocity from \cite{Rucinski:2008}; the object is flagged
  as `SB2*'.
  HIP069848 is a $\delta$~Sct star (see Appendix~A).
  (This table is available in its entirety in machine-readable form)}

\end{deluxetable}
\setlength{\tabcolsep}{6pt}

\begin{figure*}
\epsscale{1.1}
\plotone{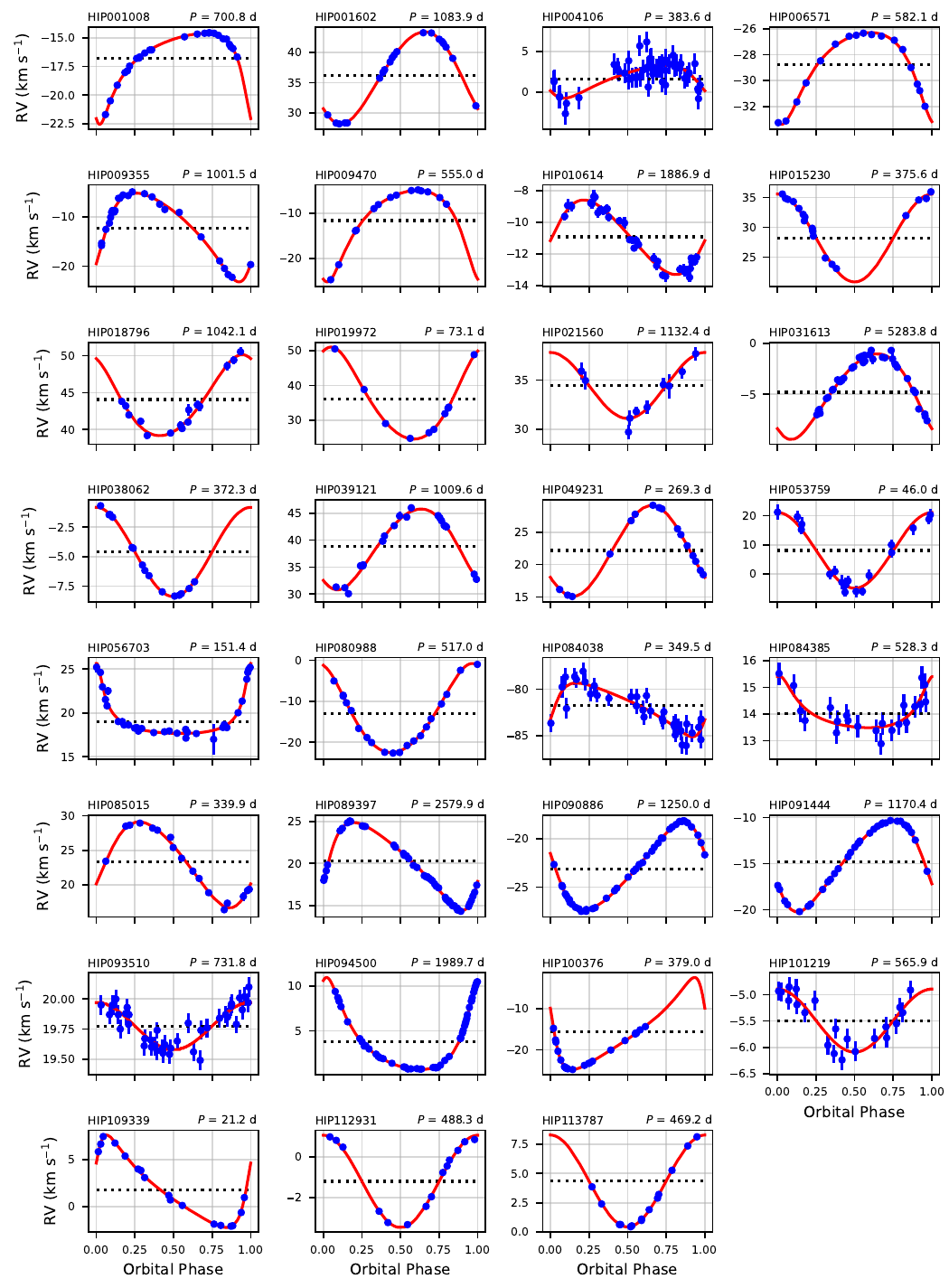}
\figcaption{Objects with single-lined spectroscopic orbits in our
sample. Orbital periods are indicated in the title line of each panel.
Dotted lines mark the centre-of-mass velocity.
\label{fig:sb1}}
\end{figure*}

More than three dozen of our targets are found to be obvious
spectroscopic binaries, and have
sufficient RV measurements and phase coverage for orbits to be derived.
Eight of them are double-lined.  These cases are
flagged with an `SB1' or `SB2' note in Table~\ref{tab:rvsummary},
for single- and double-lined systems, respectively. We show the
orbits for the single-lined systems in Figure~\ref{fig:sb1}, along
with the observations.  The corresponding orbital elements are
listed in Table~\ref{tab:sb1}. For each one, we computed also the mass
function $f(M)$, the coefficient of the minimum secondary mass
\footnote{The minimum secondary mass of an SB1 can be expressed as
$M_2 \sin i = (P/2\pi G)^{1/3} K_1 \sqrt{1-e^2} (M_1+M_2)^{2/3}$,
and the coefficient of the mass term is the part that depends only
on the orbital elements
and physical constants.\label{foot:m2sini}} $M_2 \sin i$, and the
projected semimajor axis of the primary orbit, $a_1 \sin i$. Also included
is the root-mean-square (rms) residual $\sigma$ from the orbital solution, the number of
observations, and a multiplicative scale factor $F$, which was
applied to the internal velocity errors presented in
Table~\ref{tab:rvs1} in order to achieve reduced $\chi^2$ values near
unity.

\setlength{\tabcolsep}{3pt}
\begin{deluxetable*}{lccccccccccc}
\tablewidth{0pc}
\tablecaption{Orbital Elements and Derived Properties for the Single-lined Binaries in the Sample. \label{tab:sb1}}
\tablehead{
\colhead{Name} &
\colhead{$P$} &
\colhead{$\gamma$} &
\colhead{$K_1$} &
\colhead{$e$} &
\colhead{$\omega_1$} &
\colhead{$T_0$ (BJD)} &
\colhead{$f(M)$} &
\colhead{$M_2 \sin i$} &
\colhead{$a_1 \sin i$} &
\colhead{$\sigma$ (\kms)} &
\colhead{$F$}
\\
\colhead{} &
\colhead{(days)} &
\colhead{(\kms)} &
\colhead{(\kms)} &
\colhead{} &
\colhead{(deg)} &
\colhead{(2,400,000+)} &
\colhead{($M_{\sun}$)} &
\colhead{($M_{\sun}$)} &
\colhead{($10^6$ km)} &
\colhead{$N_{\rm obs}$} &
\colhead{}
}
\startdata
HIP001008  &  700.8        &   $-$16.766\phn\phs  &  3.96   &  0.526   &  150.72        &  57156.4        &  0.00278  &  0.1406  &  32.47        &  0.066  &  0.493 \\
HIP001008  &  \phn\phn1.1  &     0.038            &  0.13   &  0.013   &  \phn\phn0.86  &  \phm{2222}2.7  &  0.00020  &  0.0033  &  \phn0.77     &     25  &        \\ [1ex]
HIP001602  & 1083.9        &    36.209            &  7.591  &  0.0874  &  131.8         &  57239.2        &  0.04857  &  0.3648  & 112.71        &  0.063  &  0.411 \\
HIP001602  & \phm{222}2.2  &    \phn0.018         &  0.028  &  0.0044  &  \phn\phn2.5   &  \phm{2222}7.5  &  0.00057  &  0.0014  & \phn\phn0.48  &     20  &        \\ [1ex]
HIP004106  &  383.6        &    1.61              &  2.03   &  0.30    &  123           &  58519          &  0.00029  &  0.066   &  10.2         &  1.202  &  3.536 \\
HIP004106  &  \phn\phn4.3  &    0.31              &  0.33   &  0.17    &  \phn36        &  \phm{222}38    &  0.00015  &  0.012   &  \phn1.8      &     49  &        
\enddata

\tablecomments{Uncertainties for the orbital elements and for the
  derived quantities are given in the second line for each system. The
  symbol $T_0$ represents a reference time of periastron passage for
  eccentric orbits, and a time of maximum primary velocity for
  circular orbits. $f(M)$ is the binary mass function.
  The column labelled $M_2 \sin i$ contains the
  coefficient of the minimum
  secondary mass, multiplying the factor $(M_1+M_2)^{2/3}$ (see
  footnote~\ref{foot:m2sini}). The last column ($F$) lists a
  multiplicative scale factor applied to the internal velocity errors
  in order to reach reduced $\chi^2$ values near unity.  (This
  table is available in its entirety in machine-readable form.)}

\end{deluxetable*}
\setlength{\tabcolsep}{6pt}

\begin{figure*}
\epsscale{1.15}
\plotone{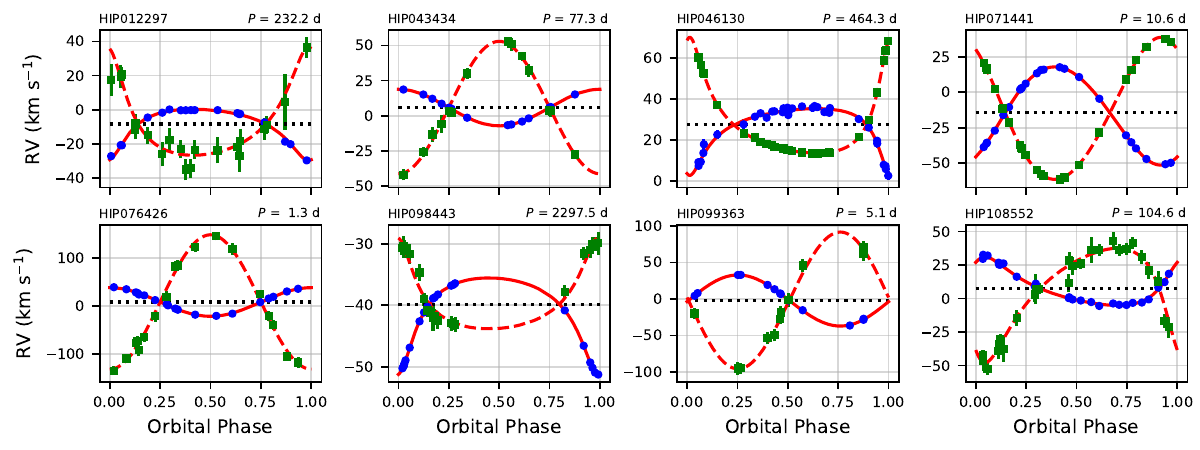}
\figcaption{Objects with double-lined spectroscopic orbits in our
sample. Orbital periods are indicated in the title line of each
panel. The solid line corresponds to the primary model, and the
dotted line marks the centre-of-mass velocity for each system. \label{fig:sb2}}
\end{figure*}

Figure~\ref{fig:sb2} displays the orbits for the double-lined binaries
in the sample. The corresponding orbital elements are collected in
Table~\ref{tab:sb2a}. Derived properties including the minimum masses,
mass ratios ($q \equiv M_2/M_1$), and projected semimajor axes are
presented separately in Table~\ref{tab:sb2b}, along with the rms
residuals for the primary and secondary, the number of observations
for each, and the scale factors for the internal errors. The
spectroscopic flux ratios we derived using TODCOR are included there
as well.

\setlength{\tabcolsep}{6pt}
\begin{deluxetable*}{lccccccccc}
\tablewidth{0pc}
\tablecaption{Orbital Elements for the Double-lined Binaries in the Sample. \label{tab:sb2a}}
\tablehead{
\colhead{Name} &
\colhead{$P$} &
\colhead{$\gamma$} &
\colhead{$K_1$} &
\colhead{$K_2$} &
\colhead{$e$} &
\colhead{$\omega_1$} &
\colhead{$T_0$ (BJD)} &
\colhead{$\Delta {\rm RV}$} &
\colhead{$\ell_2/\ell_1$}
\\
\colhead{} &
\colhead{(days)} &
\colhead{(\kms)} &
\colhead{(\kms)} &
\colhead{(\kms)} &
\colhead{} &
\colhead{(deg)} &
\colhead{(2,400,000+)} &
\colhead{(\kms)} &
\colhead{}
}
\startdata

HIP012297  &  232.23         &  $-$8.35\phs   &  15.00      &  31.6       &  0.434   &  193.8          &  56872.3          &   \nodata     &  0.0047 \\
HIP012297  &  \phn\phn0.23   &  0.15          &  \phn0.28   &  \phn2.0    &  0.013   &  \phn\phn2.0    &  \phm{2222}1.0    &   \nodata     &  0.0020 \\ [1ex]
HIP043434  &   77.323        &  5.869         &  12.986     &  47.0       &  0       &  \nodata        &  56791.022        &   \nodata     &  0.0301 \\
HIP043434  &   \phn0.025     &  0.044         &  \phn0.076  &  \phn1.4    &  \nodata &  \nodata        &  \phm{2222}0.060  &   \nodata     &  0.0028 \\ [1ex]
HIP046130  &  464.348        &  27.35         &  16.30      &  28.273     &  0.5415  &  160.97         &  57034.73         &  $-$3.18\phs &  0.0449 \\
HIP046130  &  \phn\phn0.070  &  \phn0.21      &  \phn0.32   &  \phn0.085  &  0.0021  &  \phn\phn0.34   &  \phm{2222}0.24   &   0.22       &  0.0013 
\enddata

\tablecomments{Uncertainties for the orbital elements are given on the
  second line for each system. The symbol $T_0$ represents a reference
  time of periastron passage for eccentric orbits, and a time of
  maximum primary velocity for circular orbits. $\Delta {\rm RV}$
  represents an offset between the velocity zero-points of the primary
  and secondary that we have occasionally found to be statistically
  significant, and which is caused in most cases by template mismatch.
  $\ell_2/\ell_1$ is the flux ratio at the mean wavelength of our
  observations ($\sim$5187~\AA).
  (This table is available in its entirety in machine-readable form.)}

\end{deluxetable*}
\setlength{\tabcolsep}{6pt}

\setlength{\tabcolsep}{6pt}
\begin{deluxetable*}{lccccccccc}
\tablewidth{0pc}
\tablecaption{Derived Properties for the Double-lined Binaries in the Sample. \label{tab:sb2b}}
\tablehead{
\colhead{Name} &
\colhead{$M_1 \sin^3 i$} &
\colhead{$M_2 \sin^3 i$} &
\colhead{$a_1 \sin i$} &
\colhead{$a_2 \sin i$} &
\colhead{$a_{\rm tot} \sin i$} &
\colhead{$q$} &
\colhead{$\sigma_1$ (\kms)} &
\colhead{$\sigma_2$ (\kms)} &
\colhead{$F_1$}
\\
\colhead{} &
\colhead{($M_{\sun}$)} &
\colhead{($M_{\sun}$)} &
\colhead{($10^6$ km)} &
\colhead{($10^6$ km)} &
\colhead{($R_{\sun}$)} &
\colhead{} &
\colhead{$N_1$} &
\colhead{$N_2$} &
\colhead{$F_2$}
}
\startdata

HIP012297  &  1.21   &  0.573  &  43.14      &  91.0          &  192.8        &  0.474   &  0.615  &  4.911  &  1.078 \\
HIP012297  &  0.18   &  0.053  &  \phn0.73   &  \phn5.8       &  \phn\phn8.5  &  0.031   &     21  &     20  &  1.139 \\ [1ex]
HIP043434  &  1.36   &  0.375  &  13.808     &  50.0          &  91.7         &  0.2762  &  0.140  &  2.932  &  1.127 \\
HIP043434  &  0.10   &  0.018  &  \phn0.083  &  \phn1.5       &  \phn2.1      &  0.0082  &     14  &     14  &  1.129 \\ [1ex]
HIP046130  &  1.606  &  0.926  &  87.5       &  151.77        &  343.9        &  0.577   &  1.144  &  0.197  &  1.078 \\
HIP046130  &  0.024  &  0.031  &  \phn1.7    &  \phn\phn0.32  &  \phn\phn2.5  &  0.011   &     33  &     33  &  1.077 
\enddata

\tablecomments{Uncertainties for the derived quantities are given on
  the second line for each system. $a_1$ and $a_2$ are the semimajor
  axes of the primary and secondary, while $a_{\rm tot}$ is the sum
  of the two (expressed in units of the solar radius), and $q$
  is the mass ratio $M_2/M_1$.
  The column with the $F$ headings
  contains the multiplicative scale factors applied to the internal RV
  errors for the primary and secondary, in order to achieve reduced
  $\chi^2$ values of unity. (This table is available in its entirety
  in machine-readable form.)}

\end{deluxetable*}
\setlength{\tabcolsep}{6pt}

For binaries with orbits, the mean velocities listed in
Table~\ref{tab:rvsummary} are the centre-of-mass velocities ($\gamma$)
with their associated uncertainty, taken from Tables~\ref{tab:sb1} or
\ref{tab:sb2a}. These are the proper values to use for calculating
Galactic orbits. One of our targets (HIP017878) is an overcontact
binary of the W~UMa class, for
which our velocities are not reliable. In that case, the $\gamma$
velocity listed was taken from the work of \cite{Rucinski:2008}.

\begin{figure*}
\epsscale{1.15}
\plotone{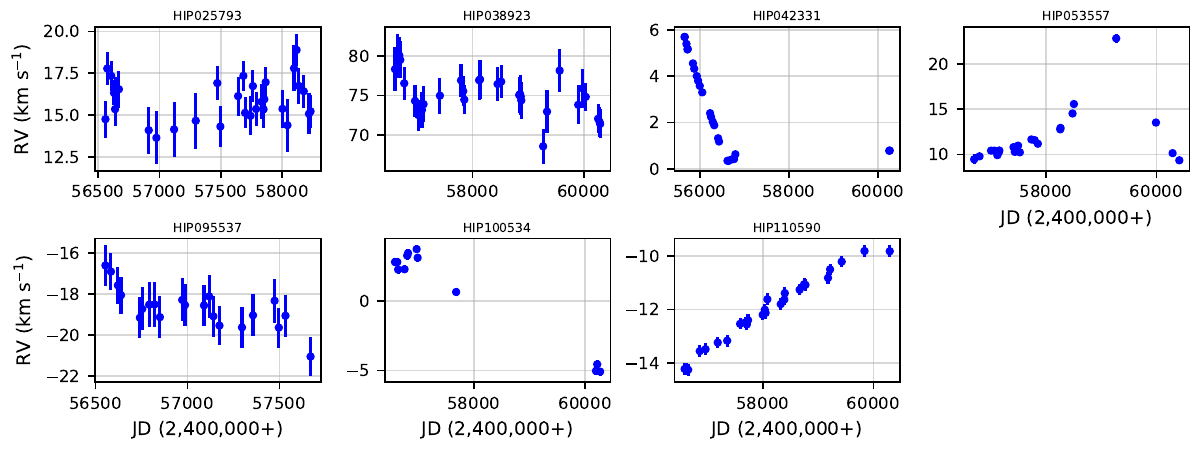}
\figcaption{Targets with obvious long-term trends in their RVs.
\label{fig:long}}
\end{figure*}

A few other objects are also obvious binaries that show trends in
their RVs, but their periods are longer than the timespan of the
observations. The clearest examples are shown graphically in Figure~\ref{fig:long},
and are flagged as `VAR' in Table~\ref{tab:rvsummary}.  Many others
have been identified as binaries or possible binaries by visual
inspection of our RVs, or by comparing our RVs with the median RVs
listed in the Gaia DR3 catalogue.\footnote{We point out that the median
Gaia velocities, which are based on about 3~yr of satellite monitoring,
were used here only to aid the detection of binaries, and did not
enter into the calculation of the mean RVs listed in
Table~\ref{tab:rvsummary}, nor into
the solution of the orbits.}
These cases are discussed individually
in Appendix~A, and are also flagged in the notes to
Table~\ref{tab:rvsummary} as `VAR' or `VAR?', depending on our degree of
confidence in their RV variability. It goes without saying that the
mean velocities listed in Table~\ref{tab:rvsummary} for the `VAR'
and `VAR?' objects are not necessarily representative of their
centre-of-mass velocities. Furthermore, in some of them the
variability may be due to pulsation rather than binarity.

For 21 of our targets, the Gaia DR3 catalogue reports `non-single star
solutions', representing the detection of binary motion, either on the
plane of the sky (from astrometry only) or perpendicular to it (RVs),
or both. In all but three of these cases, our own RVs also reveal the
binary nature of the object. The three that we missed are HIP009008,
HIP025386, and HIP033263. Full single-lined spectroscopic
orbits are presented by Gaia for the last two of these (flagged as `SB1[GAIA]'
in Table~\ref{tab:rvsummary}), while the first was identified
by Gaia as a binary from the curvature of its motion on the plane of the sky.
We indicate this astrometric detection in the table with `AST[GAIA]'.
For the last two cases, as well as for HIP062810,
the table gives the $\gamma$ velocity as determined by Gaia,
instead of the mean RV from our own measurements. We also adopted the
Gaia $\gamma$ velocity for HIP005680, which we detected as variable from our
own observations, but have insufficient data to determine an orbit.
All Gaia non-single
star solutions for the objects in our sample are gathered in a table
in Appendix~A, with an explanation of the different classes.                      

In many other cases, the Gaia DR3 catalogue or other literature
sources, such as the Washington Double Star catalogue
\citep[WDS;][]{Worley:1997, Mason:2001}, identify nearby visual
companions that are considered to be physically associated with the
target. We have flagged these cases in Table~\ref{tab:rvsummary} as
`VIS', but only if
they have not already been identified as RV variables. They are
mentioned as well in the Appendix. For
these kinds of visual binaries, the orbital periods are presumed to be
long, so the mean velocities in the table are not expected to be
biased as much as the `VAR' or `VAR?' systems might be.

Other long-period binaries can also be identified from
the difference $\Delta\mu$ between their Hipparcos and Gaia proper motions
\citep[see, e.g.,][]{Kervella:2022},
which are separated by an interval of 24.75~yr between the
mean catalogue epochs. Any such
differences are suggestive of orbital motion. To this end, we
have consulted the catalogue of astrometric accelerations of
\cite{Brandt:2021}, and identified about a dozen of our targets
with $\Delta\mu$ values larger than 3$\sigma$ in either coordinate.
All except for two of these cases are already identified as binaries
by one of the other methods above. The two that are not, HIP037104 and HIP095138, are
flagged as `ACC' in Table~\ref{tab:rvsummary}. All of these instances
are mentioned in the individual notes in the Appendix. Their mean RVs
may be expected to be little affected by their binarity, as in the
visual binaries mentioned above.

Finally, other long-period binaries are revealed by the quality of
the astrometric solutions from Gaia DR3, as quantified by the
RUWE value reported in the catalogue.
This metric is expected to be near 1.0 for sources with well-behaved
solutions, while values larger than about 1.4 are often indicative of
unmodelled binary motion, or some other problem \citep[see][]{Lindegren:2018,
Belokurov:2020}. Even sources with RUWE in the range 1.0--1.4 may also
be binaries in some cases \citep[e.g.,][]{Stassun:2021}. Of all the
objects in our sample, 41
have ${\rm RUWE} > 1.4$, but almost all of these are already recognized as
binaries in one or more of the ways explained above. The six targets
that are not, are flagged in Table~\ref{tab:rvsummary} with the code
`RUWE', and are mentioned in the Appendix.

All in all, there are 39 new spectroscopic orbital solutions derived
from our measurements in this work (31 SB1s and 8 SB2s), 4 from Gaia, and one
that was previously published (the W~UMa system HIP017878), for a total
of 44.

\section{Peculiar Velocities and Runaway Status}
\label{sec:spacevelocities}

With the mean RVs in Table~\ref{tab:rvsummary}, along with the parallaxes
and proper motions from the Gaia DR3 catalogue (Table~\ref{tab:sample}), we
proceeded to compute the peculiar velocities of our targets relative to the
Local Standard of Rest (LSR). They are listed in the last column of
Table~\ref{tab:sample}, with uncertainties that account for contributions
from all observational errors, as well as from the solar motion itself.
The components of the solar motion adopted
here are $(UVW)_{\sun} = (+11.10, +12.24, +7.25)~\kms$ \citep{Schonrich:2010},
with formal uncertainties of (0.75, 0.47, 0.37)~\kms, as reported by those authors.

\begin{figure}
\epsscale{1.18}
\plotone{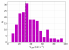}
\figcaption{Peculiar LSR velocities for our targets. The dashed line
marks the conventional 30~\kms\ dividing line above which stars are
considered runaways. Two objects in our sample are outside of the
plot with even higher velocities: HIP084038 (129~\kms),
and HIP004106 (253~\kms).
\label{fig:LSR}}
\end{figure}

Figure~\ref{fig:LSR} shows the histogram of the LSR velocities for our
sample, with the conventional 30~\kms\ threshold for runaway stars indicated
with a dashed line. We find that the majority of our stars (about 2/3) have
peculiar velocities higher than this, indicating that the original selection
of runaway candidates by \cite{Tetzlaff:2011}, carried out in the absence of RV
information at the time, was quite effective. The remaining $\sim$1/3 of
our targets have lower velocities, and are normal stars or
potential walkaway stars.

Two of our stars have LSR velocities in excess of 100~\kms: HIP084038 (129~\kms)
and HIP004106 (253~\kms). Interestingly, our RV monitoring has revealed
that both of these objects happen to be single-lined spectroscopic
binaries, with relatively long orbital periods of 349 and 383~days. They
are also both among the more distant stars in our sample, at about 670 and
880~pc, respectively.

\section{Traceback of Runaway Stars: A Pilot study}
\label{sec:traceback}

The results from the preceding sections bring a meaningful improvement
in the reliability of the RVs for our runaway targets, and provide the missing
velocities for the ones lacking the corresponding entries in the Gaia DR3 catalogue. The
impact of binarity is fully accounted for through the determination
of spectroscopic orbital solutions, where possible, or it is at least alleviated
by identifying wider binaries detectable by astrometric means,
which will presumably have less of an effect on the mean velocities.
We now turn to the use of this information and the improved space velocities
it enables.

Tracing back the 3D motion of a runaway star in the Galactic potential can
determine whether, at some time in the past, it
was sufficiently close in space to the location of the
agent that may have caused its ejection at that time, such as a young
association (dynamical ejection mechanism) or a neutron star (via a
supernova explosion). As core-collapse supernovae are expected to
occur within OB associations, in addition to tracing back a runaway
star and a neutron star as a pair, one must generally also consider
simultaneously tracing
back triplets consisting of the runaway, a neutron star, and an
association. Any encounters placing the supernova ejection outside of the
association would not be credible.

The 3D velocities of our runaway stars are now well known from the
astrometric information in the Gaia catalogue and our extensive RV
monitoring.  For the associations, their 3D motions are in general
sufficiently well established based on either Hipparcos or Gaia data
\citep[see, e.g.,][]{Neuhauser:2020, Tetzlaff:2010}. Neutron star
motions, on the other hand, are more problematic: their distances are
often poorly determined, and except for a few cases with indirect
determinations, their RVs are essentially unknown, and must be drawn
from statistical distributions inferred from their 2D motions under
certain assumptions \citep[e.g.,][]{Hobbs:2005}.

Here we have performed Monte Carlo simulations to account for all
observational uncertainties involved in the traceback, particularly
the poorly established distances and the unknown RVs of the neutron stars
\citep[see also][]{Hoogerwerf:2000, Hoogerwerf:2001}. The latter two factors
are the dominant contributors to the uncertainty. Integrating the
Galactic orbits to compare our targets against all known neutron stars
and all young associations represents a computationally
expensive effort of a scale that is beyond the scope of the present paper.
Instead, in this section we exemplify this
procedure by performing a more limited exercise, considering four of
the nearest, well known, optically visible neutron stars,
among the group known as the
Magnificent Seven \citep{Treves:2001, Zampieri:2001}. Their probable
site of origin has been studied by \cite{Tetzlaff:2010, Tetzlaff:2011,
  Tetzlaff:2012} based on Hipparcos information for runaway
  stars and associations, and their relevant properties are collected in
Table~\ref{tab:magseven}.

\setlength{\tabcolsep}{6pt}
\begin{deluxetable*}{lcccccc}
\tablewidth{0pc}
\tablecaption{Subset of `Magnificent Seven' Neutron Stars Examined Here. \label{tab:magseven}}
\tablehead{
\colhead{Name} &
\colhead{R.A.} &
\colhead{Dec.} &
\colhead{Distance} &
\colhead{$\mu_{\alpha} \cos\delta$} &
\colhead{$\mu_{\delta}$} &
\colhead{References}
\\
\colhead{} &
\colhead{(hh:mm:ss)} &
\colhead{(dd:mm:ss)} &
\colhead{(pc)} &
\colhead{(mas yr$^{-1}$)} &
\colhead{(mas yr$^{-1}$)} &
\colhead{}
}
\startdata
RX J0720.4$-$3125 & 07:20:24.96   & $-$31:25:50.1   & $357^{+167}_{-87}$ & $-93.9 \pm 1.2$\phn\phs  & $52.8 \pm 1.3$\phn    & 1, 2 \\ [0.5ex]
                  &               &                 & $278^{+222}_{-85}$ & $-92.8 \pm 1.4$\phn\phs  & $55.3 \pm 1.7$\phn    & 3    \\ [1ex]
RX J1308.6+2127   & 13:08:48.27   & +21:27:06.8     & $600 \pm 200$      & $-207 \pm 20$\phn\phs    & $84 \pm 20$       & 4, 5 \\ [1ex]
RX J1605.3+3249   & 16:05:18.52   & +32:49:18.0     & $358 \pm 58$   & $-25 \pm 16$\phs & $142 \pm 15$\phn  & 6, 7 \\ [1ex]
RX J1856.5$-$3754\tablenotemark{a} & 18:56:35.41   & $-$37:54:35.8   & $123^{+15}_{-12}$  & $325.9 \pm 1.3$\phn\phn & $-59.2 \pm 1.2$\phn\phs  & 8    \\
                  &               &                 & $161^{+17}_{-14}$  &                  &                   & 2    
\enddata
\tablecomments{References:
1) \cite{Kaplan:2005};
2) \cite{Kaplan:2007};
3) \cite{Eisenbeiss:2010};
4) \cite{Kaplan:2002};
5) \cite{Motch:2009};
6) \cite{Kaplan:2003};
7) \cite{Motch:2005};
8) \cite{Walter:2010}.}
\tablenotetext{a}{The RV is believed to be near zero, based on the
fact that the associated bow shock shows an inclination suggesting
the neutron star moves almost exactly in the plane of the
sky \citep{vanKerkwijk:2001}. \cite{Tetzlaff:2011} and
\cite{Mignani:2013} proposed that it
may have originated in the Upper Scorpius association, about 0.5~Myr ago.}
\end{deluxetable*}
\setlength{\tabcolsep}{6pt}

The traceback calculations were performed using the algorithm and
software developed for this purpose by \cite{Neuhauser:2020},
which adopts a three-component model for the Galactic potential. We ran
$3\times 10^6$ Monte Carlo simulations for each pair consisting of a
neutron star and a runaway from our sample. For pairs coming closer than
1~pc of each other, the results were analysed in more detail
\citep[see][]{Hoogerwerf:2001}.

A careful review of the results, considering the four Magnificent
Seven neutron stars in Table~\ref{tab:magseven} along with young ($< 50$~Myr) nearby OB
associations within 300 pc, revealed no credible close encounters with
any of our runaway stars, with one possible exception. The analysis
for HIP076768 and the pulsar RX\,J0720.4$-$3125 appeared initially
promising, suggesting they could have been located at the same place
at the same time within the $\beta$~Pic/Capricorn group. This would
work for either of the two published parallaxes for RX\,J0720.4$-$3125
($3.6 \pm 1.6$ mas or $2.8 \pm 0.9$ mas; see Table~\ref{tab:magseven} for
references). The encounter, as indicated in about 8000 out of $3 \times 10^6$
Monte Carlo runs, would have taken place ca.\ 0.35~Myr ago
within a few parsecs of the centre of the $\beta$~Pic group. This
flight time is shorter than the neutron star spin-down upper age
limit.  The runaway star HIP076768 is a known member of the young
Upper Scorpius association \citep[e.g.,][]{Miret-Roig:2022}), but from
its $XYZ$ position, it could also be a member of the much larger
$\beta$~Pic/Capricorn group. While its kinematics are not quite
typical for $\beta$~Pic/Capricorn, that is not unexpected for an
ejected runaway star.

However, an assessment of the credibility (or probability) of this
encounter indicated that it is not high. Specifically, tracing back
the neutron star and this young group with their actual properties, but the
runaway star HIP076768 with a random direction of motion, we found
encounters with the same neutron star within the same group in a
similar number of simulations as above ($\sim$10,000 out of $3 \times
10^6$ runs). Furthermore, HIP076768 is a known visual binary, as
mentioned in the Appendix, and wide binaries such as this are less likely to be
ejected as runaway stars and to remain bound. Therefore, we consider the
likelihood of a true encounter between HIP076768 and RX\,J0720.4$-$3125
within the $\beta$~Pic/Capricorn group to be low.

While the null result
from this limited test may be disappointing, a more comprehensive
analysis is planned in which our list of targets will be paired
against several hundred neutron stars with sufficiently precise input data.
The outcome will be reported in a follow-up publication.

\section{Concluding Remarks}
\label{sec:conclusions}

Runaway stars have received a good deal of attention in the last
decade or two, and the question of which of the formation mechanisms
dominates the creation of these objects has been a subject of much
debate \citep[see, e.g.,][]{Hoogerwerf:2000, Fujii:2011, McEvoy:2017,
Evans:2020, Bhat:2022}. Runaway stars are of interest for
a number of reasons.  For example, if tracing back the Galactic
paths of a runaway star and a neutron star reveals a close encounter in
space and in time, this could be evidence that the supernova may have
exploded then and there. In that case, the flight time of the neutron
star (i.e., its kinematic age), along with the age of the parent
association (on the assumption of contemporaneous star formation),
can provide an estimate of the progenitor's mass, and other properties,
with the aid of stellar evolution models.
Runaway stars have also been applied to investigate
the source of the radioactive
$^{60}$Fe deposited in the Earth's crust and ocean sediments by
recent, nearby, core-collapse supernovae \citep[see, e.g.,][]{Fry:2015,
Hyde:2018}. Tantalizing evidence of
the location and time of at least one of perhaps several such
events was reported by \cite{Neuhauser:2020}. It involved the O-type runaway
star $\zeta$~Oph and the radio pulsar PSR~1706$-$16, split apart and ejected
at high velocities some 1.78~Myr ago, from distance of 107~pc.

Much of the work on runaways has focused on OB stars. There are
good reasons for this: they
are brighter, a significant fraction of them show high space
velocities, their youth makes them more favourable candidates, and
they appear more frequently as former binary companions of the
even more massive supernova progenitors. Many lists of potential OB
runaway stars have been followed up over the years
\citep[e.g.,][]{Hoogerwerf:2001, Mdzinarishvili:2004, Dincel:2015,
  Maiz-Apellaniz:2018, Carretero:2023, Guo:2024}. Later-type
examples have been considered as well, including the works of
\cite{Tetzlaff:2011}, \cite{Lux:2021}, \cite{Teklehaimanot:2024}, and
others. The present project draws from the candidate list released in
the first of these later-type studies.

Our more than decade-long spectroscopic monitoring program, augmented
with earlier archival observations, was
designed to determine accurate mean RVs for nearly 190 runaway (or
walkaway) candidates of spectral types A and later, accounting for
their binary nature where needed. For the most part, these objects
had no RV measurements at the time the \cite{Tetzlaff:2011} list was
released. As a result of this work, fully characterized orbits
are now available for 44 spectroscopic binaries among our targets (39 of them new),
both single- and double-lined, with orbital periods ranging between
0.3 and 5300 days. Only a fraction of them were reported in the Gaia DR3
list of non-single-star solutions.
Many other targets have been identified here as being binary or multiple systems,
based on astrometric information from the WDS, Gaia, or other sources.
In total, about 52 per cent of the stars show either spectroscopic and/or
astrometric evidence of multiplicity. In another 10 per cent of the cases,
there are hints of velocity variability that need to be confirmed. For
the remaining 38 per cent, no evidence of binarity is apparent.

The combination of the highly precise Gaia DR3 parallaxes and proper
motions for these objects, along with the new RV information we have
obtained, allows for much more accurate space velocities to be
calculated for this sample than previously possible. This is an
essential ingredient for investigating the time and place of
their origin in the Galaxy.
Most of our targets (about 2/3) have velocities relative to the LSR that
are above 30~\kms, the classical threshold for considering an
object a runaway star. The most extreme is HIP004106, moving at
$V_{\rm LSR} = 253 \pm 5$~\kms.

As a prelude to a larger investigation, here we have carried out a
limited modelling exercise in which we traced back the orbits of our targets
along with those of four well-known neutron stars from the
Magnificent Seven group. We found no credible
cases in which a star from our sample and a neutron star were sufficiently near
each other in the recent past.

\begin{acknowledgements}

A large number of observers and colleagues at the CfA have contributed
to gathering the spectroscopic observations for this project over the years.
We thank them all.
We are also grateful to Robert Davis and Jessica Mink for maintaining the
Digital Speedometer and TRES databases at the CfA,
and to M.\ Mugrauer (U Jena) for alerting us to an error in the
original peculiar velocities of our targets.
We thank Frank Gie\ss ler (U Jena), whose software
was used in this paper for tracing back neutron stars,
runaway stars, and associations.
RN and SAH would also like to thank Baha Din\c{c}el,
Surodeep Sheth, and Luca Cortese for support in the traceback
calculations. We additionally acknowledge helpful comments from the
anonymous referee.
VVH and RN would like to thank Deutsche Forschungsgemeinschaft (DFG)
for financial support through grant numbers NE~515/61-1 and 65-1.
VVH acknowledges support from the Yerevan State University in the 
framework of an internal grant.
This research
has made extensive use of the SIMBAD and VizieR databases, operated at
the CDS, Strasbourg, France, and of NASA's Astrophysics Data System
Abstract Service.  We also acknowledge the use of the Washington
Double Star Catalogue, maintained at the U.S. Naval Observatory. The
work has used data from the European Space Agency (ESA) mission {\it
  Gaia} (\burl{https://www.cosmos.esa.int/gaia}), processed by the {\it
  Gaia} Data Processing and Analysis Consortium (DPAC,
\burl{https://www.cosmos.esa.int/web/gaia/dpac/consortium}). Funding
for the DPAC has been provided by national institutions, in particular
the institutions participating in the {\it Gaia} Multilateral
Agreement.

\end{acknowledgements}

\section{Data Availability}

The data underlying this article are available in the article and in its online supplementary material.


\section{Appendix A}
\label{sec:appendix}

We include below notes for many of our targets, pertaining to their
binarity or other details of interest. In 21 cases the Gaia DR3
catalogue has reported non-single star models of various kinds. They are
defined as follows:
\vskip 5pt
\noindent{$\bullet$~`Orbital' = Orbital model for an astrometric binary};
\vskip 5pt
\noindent{$\bullet$~`SB1' = Single-lined spectroscopic binary model};
\vskip 5pt
\noindent{$\bullet$~`AstroSpectroSB1' = Combined astrometric + single-lined spectroscopic orbital model};
\vskip 5pt
\noindent{$\bullet$~`Acceleration7' = Acceleration model with 7 parameters (astrometry only)};
\vskip 5pt
\noindent{$\bullet$~`Acceleration9' = Acceleration model with 9 parameters (astrometry only)};
\vskip 5pt
\noindent{$\bullet$~`SecondDegreeTrendSB1' = Second degree polynomial fit to the RV trend}.
\vskip 5pt
The parameters of these Gaia solutions for our targets are collected
in Table~\ref{tab:gaiaelem}, for the benefit of the reader. In another
dozen cases, the catalogue of astrometric accelerations of \cite{Brandt:2021}
indicates they are long-period binaries.

\setlength{\tabcolsep}{4pt}
\begin{deluxetable*}{llccccccccccc}
\tablewidth{0pc}
\tablecaption{Gaia DR3 Non-Single Star Models for our Targets. \label{tab:gaiaelem}}
\tablehead{
\colhead{Name} &
\colhead{Gaia model} &
\colhead{CfA} &
\colhead{$P$} &
\colhead{$T_{\rm peri}$} &
\colhead{$e$} &
\colhead{$K_1$} &
\colhead{$\gamma$} &
\colhead{$a_{\rm phot}$} &
\colhead{$i$} &
\colhead{$\omega_2$} &
\colhead{$\Omega$} &
\colhead{$\omega_1$}
\\
\colhead{} &
\colhead{} &
\colhead{} &
\colhead{(day)} &
\colhead{(BJD)} &
\colhead{} &
\colhead{(\kms)} &
\colhead{(\kms)} &
\colhead{(mas)} &
\colhead{(deg)} &
\colhead{(deg)} &
\colhead{(deg)} &
\colhead{(deg)}
}
\startdata
%
HIP001008  &  AstroSpectroSB1 &  SB1     &   691.6        &   57168.5        &  0.511  &   \nodata   & $-$16.956\phs  &  0.841    &  149.8       &  163.1       &  142.2       &  154.5        \\
HIP001008  &                  &          &   \phn\phn3.1  &   \phm{2222}3.8  &  0.022  &   \nodata   &  \phn0.040     &  0.023    &  \phn\phn3.7 &  \phn\phn8.4 &  \phn\phn7.7 &  \phn\phn2.0  \\ [1.5ex]
HIP001602  &  Orbital         &  SB1     &   926          &   57968          &  0.185  &   \nodata   &  \nodata       &  2.150    &  48.9        &  209.5       &  143.2       &  \nodata      \\
HIP001602  &                  &          &   \phn17       &   \phm{222}16    &  0.039  &   \nodata   &  \nodata       &  0.022    &  \phn2.4     &  \phn\phn7.8 &  \phn\phn2.5 &  \nodata      \\ [1.5ex]
HIP005680  &  AstroSpectroSB1 &  VAR     &   9388         &   57452.3        &  0.850  &   \nodata   & $-$28.75\phs   &  7.7      &  70.57       &  254.6       &  7.8         &  73.21        \\
HIP005680  &                  &          &   1835         &   \phm{2222}1.6  &  0.021  &   \nodata   &  \phn0.14      &  1.1      &  \phn0.78    &  \phn\phn1.5 &  1.9         &  \phn0.81     \\ [1.5ex]
HIP009008  &  Acceleration7   &  \nodata &   \nodata      &   \nodata        & \nodata &   \nodata   &  \nodata       &  \nodata  &  \nodata     &  \nodata     &  \nodata     &  \nodata      \\ [1.5ex]
HIP009355  &  Orbital         &  SB1     &   929          &   57587          &  0.51   &   \nodata   &  \nodata       &  1.29     &  67.6        &  245.5       &  140.8       &  \nodata      \\
HIP009355  &                  &          &   \phn85       &   \phm{222}27    &  0.15   &   \nodata   &  \nodata       &  0.17     &  \phn4.4     &  \phn\phn1.6 &  \phn\phn2.5 &  \nodata      \\ [1.5ex]
HIP019972  &  SB1             &  SB1     &   73.203       &   57384.7        &  0.137  &   13.59     &  37.95         &  \nodata  &  \nodata     &  \nodata     &  \nodata     &  355.1        \\
HIP019972  &                  &          &   \phn0.052    &   \phm{2222}1.5  &  0.017  &   \phn0.20  &  \phn0.19      &  \nodata  &  \nodata     &  \nodata     &  \nodata     &  \phn\phn7.2  \\ [1.5ex]
HIP021560  &  SB1             &  SB1     &   1016         &   57762          &  0.29   &   3.83      &  32.33         &  \nodata  &  \nodata     &  \nodata     &  \nodata     &  227          \\
HIP021560  &                  &          &   \phn142      &   \phm{222}98    &  0.11   &   0.48      &  \phn0.43      &  \nodata  &  \nodata     &  \nodata     &  \nodata     &  \phn43       \\ [1.5ex]
HIP025386  &  SB1             &  \nodata &   555          &   57414          &  0.235  &   0.985     & $-$48.000\phs  &  \nodata  &  \nodata     &  \nodata     &  \nodata     &  251          \\
HIP025386  &                  &          &   \phn12       &   \phm{222}26    &  0.098  &   0.063     &  \phn0.043     &  \nodata  &  \nodata     &  \nodata     &  \nodata     &  \phn19       \\ [1.5ex]
HIP033263  &  SB1             &  \nodata &   395.8        &   57286          &  0.51   &   0.410     &  4.692         &  \nodata  &  \nodata     &  \nodata     &  \nodata     &  115          \\
HIP033263  &                  &          &   \phn\phn7.7  &   \phm{222}11    &  0.11   &   0.054     &  0.037         &  \nodata  &  \nodata     &  \nodata     &  \nodata     &  \phn18       \\ [1.5ex]
HIP042331  &  Acceleration7   &  VAR     &   \nodata      &   \nodata        & \nodata &   \nodata   &  \nodata       &  \nodata  &  \nodata     &  \nodata     &  \nodata     &  \nodata      \\
HIP042331  &  SecondDegreeTrendSB1 & VAR &   \nodata      &   \nodata        & \nodata &   \nodata   & $4.4 \pm 1.1$  &  \nodata  &  \nodata     &  \nodata     &  \nodata     &  \nodata      \\ [1.5ex]
HIP062810  &  SB1             &  \nodata &   812          &   57672          &  0.251  &   7.01      & $-$33.99\phs   &  \nodata  &  \nodata     &  \nodata     &  \nodata     &  81           \\
HIP062810  &                  &          &   \phn15       &   \phm{222}47    &  0.098  &   0.87      &  \phn0.41      &  \nodata  &  \nodata     &  \nodata     &  \nodata     &  21           \\ [1.5ex]
HIP076426  &  SB1             &  SB2     &   1.339549     &   57388.33       &  0.022  &   30.73     &  7.58          &  \nodata  &  \nodata     &  \nodata     &  \nodata     &  33           \\
HIP076426  &                  &          &   0.000026     &   \phm{2222}0.24 &  0.022  &   \phn0.67  &  0.50          &  \nodata  &  \nodata     &  \nodata     &  \nodata     &  64           \\ [1.5ex]
HIP080988  &  Orbital         &  SB1     &   511.39       &   57411.2        &  0.085  &   \nodata   &  \nodata       &  3.882    &  96.90       &  181.5       &  18.32       &  \nodata      \\
HIP080988  &                  &          &   \phn\phn0.73 &   \phm{2222}8.0  &  0.012  &   \nodata   &  \nodata       &  0.015    &  \phn0.50    &  \phn\phn5.4 &  \phn0.40    &  \nodata      \\ [1.5ex]
HIP085015  &  SB1             &  SB1     &   336.5        &   57307          &  0.08  &    6.41      &  23.50         &  \nodata  &  \nodata     &  \nodata     &  \nodata     &  11           \\
HIP085015  &                  &          &   \phn\phn8.7  &   \phm{222}34    &  0.11  &    0.35      &  \phn0.56      &  \nodata  &  \nodata     &  \nodata     &  \nodata     &  39           \\ [1.5ex]

HIP090886  &  Acceleration7   &  SB1     &   \nodata      &   \nodata        & \nodata &   \nodata   &  \nodata       &  \nodata  &  \nodata     &  \nodata     &  \nodata     &  \nodata      \\ [1.5ex]
HIP091444  &  Acceleration7   &  SB1     &   \nodata      &   \nodata        & \nodata &   \nodata   &  \nodata       &  \nodata  &  \nodata     &  \nodata     &  \nodata     &  \nodata      \\ [1.5ex]
HIP097957  &  Acceleration9   &  VAR     &   \nodata      &   \nodata        & \nodata &   \nodata   &  \nodata       &  \nodata  &  \nodata     &  \nodata     &  \nodata     &  \nodata      \\ [1.5ex]
HIP098443  &  Acceleration7   &  SB2     &   \nodata      &   \nodata        & \nodata &   \nodata   &  \nodata       &  \nodata  &  \nodata     &  \nodata     &  \nodata     &  \nodata      \\ [1.5ex]
HIP112272  &  Acceleration7   &  VAR     &   \nodata      &   \nodata        & \nodata &   \nodata   &  \nodata       &  \nodata  &  \nodata     &  \nodata     &  \nodata     &  \nodata      \\
HIP112272  &  SecondDegreeTrendSB1 & VAR &   \nodata      &   \nodata        & \nodata &   \nodata   & $35.4 \pm 1.2$ &  \nodata  &  \nodata     &  \nodata     &  \nodata     &  \nodata      \\ [1.5ex]
HIP112931  &  SB1             &  SB1     &   480.2        &   57385          &  0.055  &   2.284     & $-$1.248\phs   &  \nodata  &  \nodata     &  \nodata     &  \nodata     &  189          \\
HIP112931  &                  &          &   \phn\phn3.5  &   \phm{222}24    &  0.026  &   0.049     &  0.037         &  \nodata  &  \nodata     &  \nodata     &  \nodata     &  \phn17       \\ [1.5ex]
HIP113787  &  Orbital         &  SB1     &   492.7        &   57428          &  0.18   &   \nodata   &  \nodata       &  0.562    &  53.2        &  12.9        &  173.7       &  \nodata      \\
HIP113787  &                  &          &   \phn\phn8.2  &   \phm{222}39    &  0.13   &   \nodata   &  \nodata       &  0.036    &  \phn7.6     &  \phn3.4     &  \phn\phn6.5 &  \nodata      
\enddata
\tablecomments{The second line for most objects contains the uncertainties
of the orbital elements, as reported by Gaia. The third column (CfA)
indicates our conclusion regarding binarity,
based solely on our own RVs. The elements in the table have their usual meaning.
Times of periastron passage, $T_{\rm peri}$, are referenced to JD 2,400,000.
Column $a_{\rm phot}$ represents the angular semimajor axis of the photocentre,
as Gaia generally does not resolve the components.
Column $\omega_2$ gives the argument of periastron for the secondary in the
astrometric orbit,
while $\omega_1$ is the angle for the primary obtained using only the RVs.}
\end{deluxetable*}
\setlength{\tabcolsep}{6pt}

\vskip 10pt
\noindent{\bf HIP001008}: SB1 with $P = 700.8$~d. Gaia DR3
`AstroSpectroSB1' solution.

\napp{\bf HIP001602}: SB1 with $P = 1084$~d. Gaia DR3 `Orbital'
solution.

\napp{\bf HIP001733}: Two of our velocities are considerably
lower than the rest, but we have not been able to determine a period.
Possible binary.

\napp{\bf HIP002838}: Gaia DR3 lists a common proper motion
companion with the same parallax, at a separation of 40\farcs9 and
2.3~mag fainter in $G$. The Gaia RV for this companion is the same as
that of the primary, confirming the association.

\napp{\bf HIP004106}: Tentative SB1 with $P = 384$~d. This is a long-period
photometric variable (CO~Cet), classified by Hipparcos as semiregular,
and claimed to have a possible photometric period of 73.5~d. The RV variations have a low
amplitude, and seem less regular after mid-2018. They may be caused by pulsations rather than orbital motion.

\napp{\bf HIP005680}: The RVs clearly indicate this is a binary, possibly with
a rather eccentric orbit. Gaia DR3 reports an
`AstroSpectroSB1' solution with a very long but uncertain period,
and a high eccentricity. We adopt the $\gamma$ velocity from Gaia as the
best representation of its mean velocity.
There is a much wider companion ($\rho =
33\farcs6$, $\Delta G = 4.2$~mag) listed by Gaia that shares the same
parallax and proper motion. It has a median Gaia RV of $-4.04 \pm
21.89~\kms$, where the large formal error is a reflection of the
scatter, and indicates that it too may be
variable. This would make it a quadruple system. The target displays
astrometric acceleration \citep{Brandt:2021}.

\napp{\bf HIP009008}: There is only a hint of variability in our RVs.
Gaia DR3 `Acceleration7' solution.

\napp{\bf HIP009355}: SB1 with $P =1002$~d. Gaia DR3 `Orbital'
solution.  Long-period photometric variable (DE~Psc). There is a claim
of a (noisy) detection of a binary companion in a lunar occultation
observation on Christmas day, 1982, by \cite{Beskin:1987} (projected
separation $8.3 \pm 0.7$~mas, vector angle 34\arcdeg, $\Delta R = 1.3
\pm 0.1$~mag).  However, it was not seen in a subsequent lunar
occultation event by \cite{Dyachenko:2018}, who only reported a
measure of the angular diameter of the star.

\napp{\bf HIP009470}: SB1 with $P = 555$~d. No companions were
detected in a lunar occultation observation by \cite{Eitter:1979}.

\napp{\bf HIP011242}: Although our three RV measurements over a
two month period show no change, this appears to be a binary. A lunar
occultation observation by \cite{Evans:1985} detected a companion with
a projected separation of $23.0 \pm 1.0$~mas, vector angle 191\fdg6,
and $\Delta m = 1.99 \pm 0.19$~mag in a red filter. It was unresolved
in a speckle observation by \cite{Mason:1996}.

\napp{\bf HIP011339}: Clearly a binary, based on the large
difference between the median Gaia RV and ours. Our RVs alone show
possibly signs of variability, with a hint of asymmetry in the CCFs
due perhaps to blended lines from a companion. The target displays
astrometric acceleration. Gaia classifies it as a
photometric variable with short-period oscillations similar to the
$\delta$~Sct stars.

\napp{\bf HIP011663}: The median Gaia RV is slightly lower than
our three measurements obtained about two years earlier, suggesting this
may be a binary.

\napp{\bf HIP012297}: This is a binary with a 232~d period, for
which we present an SB2 orbit. The detection of the faint M-dwarf
secondary at a flux ratio $\ell_2/\ell_1$ of just 0.5 per cent is somewhat
tentative.

\napp{\bf HIP013284}: The median Gaia RV is marginally lower than
ours.  Possible binary.

\napp{\bf HIP015373}: Candidate massive member of the Psc-Eri
stream \citep{Curtis:2019}. Probable binary based on a median Gaia RV
about 2~\kms\ higher than ours.

\napp{\bf HIP016608}: The observations indicate this is a binary.
One of our RVs is 10~\kms\ lower than the rest, while the median Gaia RV
is 10~\kms\ higher. Gaia classifies it as a photometric variable with
short-period oscillations similar to the $\delta$~Sct stars.

\napp{\bf HIP016615}: The velocity is variable, indicating it is
a binary.

\napp{\bf HIP017635}: The median Gaia RV is 2~\kms\ higher than
ours, which were obtained two years earlier.
The target displays astrometric acceleration, revealing its binary nature.

\napp{\bf HIP017878}: This is a W~UMa overcontact eclipsing binary
(V1128~Tau) with a period of 0.305371~d \citep[e.g.,][]{Caliskan:2014}.
Our RVs are not reliable
because of the rapid rotation and resulting severe line blending, so
we do not report them. However, \cite{Rucinski:2008} has measured
velocities and reported a
double-lined spectroscopic orbit with a centre-of-mass velocity of
$-12.27 \pm 0.76~\kms$. We list this value in Table~\ref{tab:rvsummary}.
There is a 12\farcs1 companion (HIP017876)
about 1.1~mag fainter in $G$ that shares the parallax and proper
motion of the primary, and has a median Gaia RV of $-10.72 \pm
0.32~\kms$. This is therefore a hierarchical triple system.

\napp{\bf HIP019972}: SB1 with a period of 73~d. Gaia DR3 also
reports an `SB1' solution with the same period.

\napp{\bf HIP020513}: This is a long-period photometric variable
known as V1142~Tau.  There appears to be a roughly 2~yr periodicity in
our RVs, which may well be due to semiregular pulsation rather
than orbital motion.

\napp{\bf HIP021560}: SB1 with a period of 1147~d. Gaia DR3
reports an `SB1' solution with a similar period.

\napp{\bf HIP022104}: The Gaia DR3 catalogue lists a companion with
the same parallax and proper motion at a separation of 31\farcs5 and
4.1~mag fainter in $G$. Its median RV as reported by Gaia ($-14.03 \pm
1.10~\kms$) agrees well with our average primary velocity, supporting
the physical association.

\napp{\bf HIP023933}: Probable binary: the median Gaia RV is some
2~\kms\ higher than ours.

\napp{\bf HIP024478}: The Gaia catalogue lists a companion 4\farcs4
away that is 8.2~mag fainter in $G$, and shares the parallax and
proper motion of the primary.

\napp{\bf HIP024780}: The Gaia catalogue lists a companion 20\farcs7
away that is 8.1~mag fainter in $G$, and shares the parallax and
proper motion of the primary. The primary itself appears to have a
constant RV, within the uncertainties.
The target displays astrometric acceleration.

\napp{\bf HIP025386}: Gaia DR3 reports an `SB1' solution with a
555~d period and a small velocity semiamplitude.
The handful of RVs we obtained are not inconsistent with that orbit.
We adopt the $\gamma$ velocity from Gaia as the 
best representation of its mean velocity.

\napp{\bf HIP025793}: Herbig Ae/Be star
\citep[e.g.,][]{Thomas:2023}.  There is a hint of a long period in our
RV measurements. Gaia DR3 does not report any RV information.

\napp{\bf HIP027380}: The median Gaia RV is about 1~\kms\ larger
than ours, possibly indicating binarity. The WDS lists a visual
companion at 7\farcs3, some 5 or 6~mag fainter than the primary. Gaia
confirms the physical association from the similar proper motions and
parallaxes. The RV changes may be due to this companion.

\napp{\bf HIP027634}: Visual binary, according to the WDS, with a
3\farcs8 companion about 1.3~mag fainter. Gaia confirms the physical
association of the two stars. The primary has a constant RV.

\napp{\bf HIP028539}: The Gaia DR3 catalogue reports a large RUWE value
of 2.020, suggesting it may be a binary.

\napp{\bf HIP029639}: The median Gaia RV is 1.5~\kms\ lower than
ours, suggesting it may be a binary.

\napp{\bf HIP031498}: Very rapid rotator. This makes our RVs very
poor. Nevertheless, there is a hint of variability based on the
pattern of the measurements. Gaia DR3 reports
a median RV of $-11.58 \pm 2.11~\kms$ based on 11 measurements. 
This may be a binary.

\napp{\bf HIP031613}: Clearly a binary. Our SB1 solution with
a period of roughly 5300~d is preliminary, as the observations
do not yet cover a full cycle.

\napp{\bf HIP031807}: The median Gaia RV is 1~\kms\ lower than
ours, suggesting it may be a binary.

\napp{\bf HIP033263}: Gaia DR3 reports an `SB1' solution with a
very small semiamplitude of only $K_1 = 0.410 \pm 0.054~\kms$. Our 4
RVs show no significant change, but are not inconsistent with the Gaia
orbit. We adopt the $\gamma$ velocity from Gaia as the 
best representation of its mean velocity.

\napp{\bf HIP034729}: Other RVs from the literature, as well as
Gaia, agree with ours and indicate a constant RV. The WDS lists a
visual companion at 10\farcs8, more than 7~mag fainter than the
primary. However, the information from Gaia DR3 indicates it is
unrelated.

\napp{\bf HIP036158}: The RVs show an upward drift,
and the median Gaia RV at an
intermediate epoch is consistent with the trend. This is a binary.

\napp{\bf HIP037104}: Visual binary, according to the WDS, with a
12\farcs2 companion about 5~mag fainter. While the parallaxes are
somewhat similar, the proper motions are not, according to Gaia, so
the two stars do not appear to be associated. However, the target displays
astrometric acceleration, suggesting there may be another companion.

\napp{\bf HIP038062}: SB1 system with a 372~d period.
The WDS reports a companion at 9\farcs3
about 4.4~mag fainter, although a note says there is doubt as to its
identification. The Gaia DR3 catalogue shows no companions at this
distance.

\napp{\bf HIP038923}: Long-period photometric variable known as
V407~Pup. The RVs may be variable as well, possibly due to
semiregular pulsations.

\napp{\bf HIP039121}: We report this as an SB1 system with a
period of 1010~d. The scatter from the orbital fit is unexpectedly
small for an object with a rotational broadening as large as we
estimate ($\sim$70~\kms). We suspect it to be
double-lined, consistent with its sizeable minimum secondary mass,
but with a line blending too severe to allow us to separate the components.
In that case, the significant line broadening is the result of the blending,
and the formal individual RV uncertainties listed in Table~\ref{tab:rvs1}
may be overestimated. Gaia DR3 reports a visual companion at 34\farcs5 and
about 10.3~mag fainter in $G$, which shares the same parallax and proper motion.
Interestingly, it is bluer than the target, indicating it is likely a
white dwarf.

\napp{\bf HIP042331}: Our RVs indicate this is a binary. Gaia DR3
reports an `Acceleration7' solution, as well as a
`SecondDegreeTrendSB1' solution describing a non-linear RV
variation. Astrometric acceleration is also reported by \cite{Brandt:2021}.
While our observations alone are insufficient to determine the
period, it is possible to derive a very tentative orbital solution by
making use of the median Gaia RV. The period of such a solution is
about 3300~d, and the centre-of-mass velocity is 4.5~\kms. The
latter is similar to the value reported by Gaia (Table~\ref{tab:gaiaelem}).

\napp{\bf HIP044580}: Our 3 RV measurements show no change, but
only span a few weeks. The median Gaia RV over more than 2~yr is
consistent with our average. The Gaia catalogue reports that about a
quarter of the scans display more than one peak, suggesting a
partially resolved (sub-arcsecond) binary. A speckle observation by
\cite{Mason:1999} revealed no companions.

\napp{\bf HIP046130}: SB2 binary with $P = 464$~d. We find a significant
primary/secondary velocity offset that is likely due to template
mismatch. The primary is a rapid rotator.

\napp{\bf HIP047155}: The Gaia DR3 catalogue reports a large
RUWE value of 3.268, suggesting it may be a binary.

\napp{\bf HIP050417}: Our RVs show no significant change within
the errors, but the median Gaia RV is about 2.5~\kms\ higher than
ours, possibly due to orbital motion.

\napp{\bf HIP050719}: No significant RV change in our 3
measurements.  The median Gaia RV is 1~\kms\ higher than ours,
possibly due to a companion.

\napp{\bf HIP050999}: The median Gaia RV is nearly 1~\kms\ lower
than our 3 measurements.

\napp{\bf HIP053557}: This is clearly a binary with a period of
the order of 4000 or 5000~d. The spectrum at the maximum velocity
is double-lined, and some of the others may also be, but we are
unable to measure separate velocities for the two components in
most of the spectra.

\napp{\bf HIP053759}: SB1 binary with a period of 46~d. Gaia lists
a companion 43\farcs8 away and 7.3~mag fainter in $G$, with the same
parallax and proper motion. The median RV of this companion is listed
by Gaia as $14.92 \pm 5.06~\kms$. It is somewhat different from the
centre-of-mass velocity we derive for the primary ($\gamma = 8.05 \pm
0.48~\kms$), but the large formal error for the companion suggests it
too may be a binary. If so, this would be a hierarchical quadruple
system.

\napp{\bf HIP056703}: This SB1 with a period of 152~d is the
fainter secondary component of the visual binary ADS~8243
(WDS~J11376$-$1656), separation 13\farcs7, $\Delta V = 0.8$~mag. There
is no parallax or proper motion listed for the primary in the Gaia DR3
catalogue. However, the very small relative motion between the stars
recorded in the WDS over nearly 200~yr strongly suggests they are
physically associated. This would then be a triple system. The Gaia
catalogue reports that the SB1 is partially resolved in about half of
the observations.

\napp{\bf HIP056777}: The Gaia DR3 catalogue indicates there is a
faint companion ($\Delta G = 8.0$~mag) at a separation of 2\farcs5,
which has the same parallax and proper motion as the primary. Our RVs
for the primary show no change.

\napp{\bf HIP057241}: The median Gaia RV is formally about
2~\kms\ lower than ours, although this is only marginally significant
given the uncertainties. Gaia classifies it as a photometric variable.

\napp{\bf HIP058028}: The WDS lists a companion at 15\farcs2
nearly 10~mag fainter.  Gaia DR3 confirms it is physically associated,
and that it is a very red star.

\napp{\bf HIP058179}: The median Gaia RV is 12~\kms\ lower than
ours, indicating this is a binary.

\napp{\bf HIP058861}: This is the primary of a 30\arcsec\ visual
binary, with $\Delta V \sim 2$~mag. Gaia DR3 confirms the two objects
are physically associated.  It is a very rapid rotator. The median
Gaia RV for the primary is more than 10~\kms\ higher than ours,
indicating it is a binary. This therefore appears to be a hierarchical
triple system.

\napp{\bf HIP061018}: A planet with a 20.8~d period has been
identified orbiting this star \citep{Livingston:2018}.
The Gaia~DR3 catalogue reports a large RUWE value of 1.474,
suggesting it may have a stellar companion as well.

\napp{\bf HIP062810}: Gaia DR3 reports an `SB1' solution with a
period of 812~d. Our RVs show changes, and are not inconsistent
with that orbit. We adopt the $\gamma$ velocity from Gaia as the 
best representation of its mean velocity. Gaia also indicates
there is a common proper motion companion 4.4~mag fainter in $G$,
which is 13\farcs4 away and has the same parallax.  Its median RV,
according to Gaia, is $-35.63 \pm 0.88~\kms$. This value is not far
from the centre-of-mass velocity that Gaia reports for the primary:
$-33.99 \pm 0.41~\kms$. We conclude this is a triple system.

\napp{\bf HIP066045}: Primary of a wide (48\farcs1) visual binary,
with the companion being 1~mag fainter, according to the WDS. Gaia DR3
confirms the physical association. The median Gaia RV is more than
1~\kms\ higher than ours, hinting at binarity. If this is confirmed,
it would constitute a triple system. The median Gaia RV for the
companion, according to Gaia, is $-2.21 \pm 0.19~\kms$, not far from
the average for the primary.

\napp{\bf HIP069848}: This is a $\delta$~Sct star (MX~Vir), with a
measured photometric frequency of 6.49~d$^{-1}$ \citep[$P =
  0.154$~d;][]{Barac:2022}. Although there is no evidence of binarity,
we performed an `orbital' fit simply as a means of deriving a
representative average velocity. We obtained a period virtually the
same as the photometric one, and a RV amplitude of 3.66~\kms.

\napp{\bf HIP071712}: Long-period photometric variable known as EG~Boo.

\napp{\bf HIP073216}: The RV appears to be variable, but we are
unable to identify a plausible orbital period.

\napp{\bf HIP074425}: The Gaia DR3 catalogue reports a large RUWE
value of 3.250, suggesting this may be a binary.

\napp{\bf HIP076426}: SB2 binary with a period of 1.34~d.
The secondary contributes only
0.6 per cent of the flux. This is a known RS~CVn variable
\citep[BI~CrB;][]{Norton:2007} with a similar photometric period as the
spectroscopic orbit.  Gaia DR3 has an `SB1' solution with the same
period as ours.

\napp{\bf HIP076768}: Our RVs show no change. Gaia DR3 gives no
information on the RV. The WDS has this as
a visual binary with an astrometric orbit \citep[WDS~J15405$-$1842AB,
  $P = 79.5$~yr, $a = 0\farcs6$, magnitude difference $\sim$
  1~mag;][]{Horch:2021}.\footnote{We note that Table~4 by
  \cite{Horch:2021} gave an incorrect WDS name for this system.}
Gaia resolves the two components, but only gives the parallax and
proper motion for the primary. The target displays
astrometric acceleration, most likely from the same companion.
The All Sky Automatic Survey
\citep[ASAS;][]{Kiraga:2012} lists it as a photometric (rotational)
variable, with a period of 3.546~d and an amplitude of 0.068~mag in
$V$. See Section~\ref{sec:traceback}.

\napp{\bf HIP078171}: Visual binary with a separation of about
0\farcs7, according to the WDS, partially resolved by Gaia. The
companion is 3~mag fainter. Our RVs show no change.

\napp{\bf HIP078846}: The median Gaia RV is several \kms\ larger
than our average, suggesting it may be a binary. This is a very rapid
rotator.

\napp{\bf HIP079687}: The median Gaia RV is about 5~\kms\ higher
than our average, suggesting binarity.

\napp{\bf HIP080988}: SB1 binary with a period of 517~d. Gaia DR3
reports an `Orbital' solution with a similar period. The secondary
is invisible in our spectra, and is expected to be a low-mass star
based on the Gaia inclination angle of about 97\arcdeg.

\napp{\bf HIP084038}: SB1 binary with a period of 349~d. Two RVs
from \cite{Sperauskas:2002} agree with our orbit. This is listed as a
long-period photometric variable (V940~Her). Hipparcos classifies it
as a semiregular variable, and gives a photometric period of $313 \pm
2$~d (somewhat similar to the orbital period), and an epoch of maximum
brightness of JD~2,448,576, which coincides with the phase of
maximum RV in the orbit. Gaia DR3 reports a photometric period of
about 350~d (though with a large uncertainty), again near the orbital
period.

\napp{\bf HIP084385:} Our SB1 orbit with a period of 529~d has a low RV semiamplitude,
and is somewhat tentative. SIMBAD indicates it is a long-period photometric
variable (V942~Her). This could be the reason for the velocity variations.

\napp{\bf HIP084680}: Long-period photometric variable (V818~Her)
classified as a semiregular (SRb), according to
\cite{Jerzykiewicz:1984}, with a cycle length of tens of days. Its
brightness also appears to vary on a shorter timescale of 6.75~d
\citep{Koen:2002}. The median Gaia RV is slightly different from our
average, but this may be related to the oscillations rather than to
binarity.

\napp{\bf HIP085015}: SB1 with a period of 340~d. Gaia DR3 reports
an `SB1' solution with the same period, but less precise orbital
parameters.

\napp{\bf HIP088984}: The median Gaia RV is about 1.5~\kms\ lower
than ours. This could be a binary.

\napp{\bf HIP089397:} The period of this SB1 is 2580~d.
There is a suggestive periodicity of about 580~d in the residuals,
which could be due to photometric variability, although we find
no record of that in the literature.

\napp{\bf HIP089553}: Photometric variable (NSV~24367) with a
period of 20.4~d and a 0.1~mag amplitude \citep{Burggraaff:2018}.

\napp{\bf HIP090886}: SB1 binary with a period of 1250~d. Gaia DR3
reports an `Acceleration7' solution.

\napp{\bf HIP091444}: SB1 binary with a period of 1171~d. Gaia DR3
reports an `Acceleration7' solution.

\napp{\bf HIP091828}: This is the secondary in an 11\farcs5 visual
binary (WDS~J18434+3533), according to the WDS, in which the primary
(HIP091829, not in our sample) is 1~mag brighter. Gaia DR3 confirms the physical
association. Our RVs show no change, with a mean velocity that agrees
with the median value reported for the primary by Gaia ($-8.56 \pm
0.77~\kms$).

\napp{\bf HIP092787:} One of our RVs is almost 2~\kms\ higher than
the rest, suggesting it may be a binary.
The Gaia catalogue does not report a non-single star solution,
but indicates that the object was partially resolved 16 per cent of the
time, supporting our suspicions. This is a Mira variable (V913 Aql).

\napp{\bf HIP093510}: Our SB1 orbit with a period of 731~d has a
very small RV amplitude, and consequently a small minimum secondary
mass of $M_2 \sin i = 0.00838 \pm 0.00071 (M_1+M_2)^{2/3} M_{\sun}$.
It is possible this variability is due to pulsation, rather than
binarity. Gaia DR3 lists a red companion about 4\farcs5 away and
9.6~mag fainter in $G$, which appears to be physically associated,
based on the very similar parallaxes and proper motions. This could
then be a triple system.

\napp{\bf HIP095138}: This target displays astrometric acceleration,
suggesting binarity.

\napp{\bf HIP095537}: The WDS lists this object as a close visual
binary ($\rho = 0\farcs1$, $\Delta m = 0.3$~mag). Our RVs show a
slight downward drift, possibly caused by the companion. This is also
a long-period semiregular variable (V557~Lyr, SRd).

\napp{\bf HIP097678}: Long-period photometric variable ($\Pi$~Dra,
SRd:). A period of 21.7~d has been reported \citep{Dubath:2011}.

\napp{\bf HIP097957}: Our last RV is almost 5~\kms\ lower than the
rest, gathered 3~yr earlier. The median Gaia RV agrees with the trend.
Gaia DR3 reports an `Acceleration9' solution. This is clearly a
long-period binary. The target displays astrometric acceleration.

\napp{\bf HIP098286}: The cross-correlation functions (CCFs) give
the impression of having multiple peaks (at least three), suggesting a
composite spectrum. As we do not have enough spectra to confirm this,
the RVs reported here were derived under the assumption that it is a
single rapidly-rotating star (with all CCF peaks blended together).
The median Gaia RV agrees with the average from our 7 observations.
The Gaia DR3 catalogue lists a very red companion at 13\farcs5 with the
same parallax and proper motion, which is 9~mag fainter in the $G$
band. This companion may itself be a binary, as the Gaia catalogue
indicates it was partially resolved in most scans. In that case, the
system would be at least triple, or possibly more complex. HIP098286
is also listed as a photometric variable (NSV~24947), with a possible
period of 0.17328~d \citep{Samus:2017}, perhaps a $\delta$~Sct star.
An occultation of the primary star by asteroid 248~Lameia was observed
in June of 1998, and used to estimate the size and shape of the
asteroid \citep{Fraser:1998}.

\napp{\bf HIP098443:} This is an SB2 (BD+17~4185) with a period
of 2297~d. Gaia DR3 reports an `Acceleration7' solution.
The WDS lists several very wide companions, of which the brightest
is BD+17~4187. A WDS note indicates our target is variable, and the AAVSO
lists it as type `LB' \citep{Watson:2006}, which is a slow irregular
variable (NSV~12668).

\napp{\bf HIP098762}: Our last RV is about 1~\kms\ lower than the
previous ones, obtained 3~yr earlier. The median Gaia RV seems to
follow the trend.  We consider this to be a likely binary.

\napp{\bf HIP099070}: The Gaia DR3 catalogue reports a large RUWE
value of 1.666, suggesting this may be a binary.

\napp{\bf HIP100180}: The WDS reports this is a visual binary with
a 0\farcs12 companion and a magnitude difference of 0.9~mag (Hipparcos
detection). It is also photometrically variable
\citep[NSV~25104;][]{Samus:2017}. Our RVs show no significant change.

\napp{\bf HIP100376:} Our observations do not cover the phase of
maximum RV, but the SB1 orbit with $P = 379$~d is still well determined.

\napp{\bf HIP100534}: The variable RV indicates this is a binary
with a long period. The median Gaia RV agrees with the trend.

\napp{\bf HIP100845}: The median Gaia RV is about 3~\kms\ higher
than ours, suggesting binarity. This is a very rapid rotator.

\napp{\bf HIP101219}: While we report an SB1 orbit here ($P = 566$~d), the
low-amplitude RV variability could also be due to pulsation rather
than orbital motion.

\napp{\bf HIP101796}: Very poor RVs.
The Gaia DR3 catalogue reports a large RUWE value of 2.420, suggesting
this may be a binary.

\napp{\bf HIP102641}: Visual binary, according to the WDS. The
companion is currently at 17\farcs7, according to Gaia, and is about
3.5~mag fainter. The similar parallaxes and proper motions from Gaia
DR3 confirm the physical association. This is also supported by the
median RV reported by Gaia for the companion ($-16.62 \pm 1.02~\kms$),
which agrees with the primary velocity. Our 3 RVs of the primary show
no change, but span less than a year. The companion may itself be a
binary, as it was reported by Gaia to be partially resolved in about a
quarter of the scans. In that case, it would be a triple system.

\napp{\bf HIP102804}: Photometric variable (NSV~25356) with $P =
64.9$~d and a 0.15~mag amplitude \citep{Burggraaff:2018}. The median
Gaia RV is about 1.5~\kms\ lower than ours, suggesting this could be a
binary. Gaia reports the star was partially resolved in 20 per cent of the
scans, suggesting it is a binary. Several wider companions listed in
the WDS are not physically associated, according to Gaia.

\napp{\bf HIP104444}: A lunar occultation observation by
\cite{Eitter:1974} did not detect any companions.

\napp{\bf HIP107325}: Our last RV is marginally lower than the
rest, from 3~yr earlier. The median Gaia RV is in between.
Gaia DR3 lists a 3\farcs8 companion 7.5~mag
fainter in $G$, which has a similar parallax and somewhat similar (if
uncertain) proper motion.

\napp{\bf HIP107588}: Although we see only one set of lines in our
spectra, the CCF peak has a triangular shape different from that of
other stars with the same rotational broadening. According to the WDS,
this is a visual binary with a separation of 0\farcs6 and a magnitude
difference of 1.3~mag. It was also partially resolved by Gaia. We
suspect, then, that the CCF shape is the result of a blend between a
rapidly rotating star, presumably the primary, and a more slowly
rotating secondary, with a very small velocity separation that
we cannot resolve. Gaia classifies it as a photometric variable.

\napp{\bf HIP108552}: The secondary in this SB2 system ($P = 105$~d) contributes
only about 0.5 per cent as much flux as the primary.

\napp{\bf HIP109339}: SB1 binary with a period of 21~d. The WDS lists a wide companion
at 22\farcs9 and about 1.3~mag fainter. Gaia confirms the physical
association from the similar parallaxes and proper motions, so this is
at least a triple system.

\napp{\bf HIP110590}: This is a binary. We see an almost linear
increase in the RV over 9~yr. The median Gaia RV agrees with this
trend.

\napp{\bf HIP110993}: The median Gaia RV is 8~\kms\ higher than
ours, suggesting binarity. This is a rapid rotator.

\napp{\bf HIP111003}: Gaia lists a 4\farcs2 common proper motion
companion 8.8~mag fainter in $G$, which has the same parallax as the
primary.  Our RVs for the primary span 3~yr, and are constant.

\napp{\bf HIP111522}: The RV is variable, indicating binarity.
The orbital period is likely long. The median Gaia RV agrees
with the trend.

\napp{\bf HIP112250}: Long period irregular variable (QV~Peg).
The median Gaia RV is 1.5~\kms\ lower than ours, suggesting binarity.

\napp{\bf HIP112272}: The RVs are clearly variable, likely with
a long orbital period. Gaia DR3 reports an
`Acceleration7' astrometric solution, as well as a non-linear RV
solution of type `SecondDegreeTrendSB1'. The median Gaia RV
agrees with the trend. \cite{Brandt:2021} also reported that
the target displays significant astrometric acceleration.

\napp{\bf HIP112931}: SB1 binary with a period of 488~d. Gaia DR3
also reports an `SB1' solution, with the same period.

\napp{\bf HIP113787}: SB1 binary with a 469~d period. Gaia DR3
reports an `Orbital' solution with a similar period.

\napp{\bf HIP116987}: This is a very rapid rotator. The median
Gaia RV is about 3~\kms\ higher than ours, hinting that it may be a
binary.

\end{document}